\documentstyle[aps,preprint,epsfig]{revtex}
\long\def\comment#1{\vskip 2mm\noindent\fbox{%
\vbox{\parindent=0cm{\small\em #1}}}\vskip 2mm}
\newcommand{\Label}[1]{\label{#1}}

\begin{document}
\tightenlines  
\title{Quantum Kinetic Theory III: Quantum kinetic master equation for 
strongly condensed trapped systems}
\author{C.W.~Gardiner$^{1}$ and P.~Zoller$^2$}
\address{$^1$ School of Chemical and Physical Sciences, Victoria University, 
Wellington, New Zealand}
\address{$^2$ Institut f{\"u}r Theoretische Physik,
Universit{\"a}t Innsbruck, 6020 Innsbruck, Austria}
\maketitle
\begin{abstract}
We extend quantum kinetic theory to deal with a strongly Bose-condensed atomic 
vapor in a trap.  The method assumes that the majority of the vapor is not 
condensed, and acts as a bath of heat and atoms for the condensate.  
The condensate is described by the particle number conserving Bogoliubov
method developed by one of the authors.
We derive equations which describe the fluctuations of particle number and 
phase, and the growth of the Bose-Einstein condensate.  The equilibrium state 
of the condensate is a mixture of states with different numbers of particles 
and  quasiparticles.  It is {\em not} a quantum superposition of states with 
different numbers of particles---nevertheless, the stationary state exhibits 
the 
property of off-diagonal long range order, to the extent that this concept 
makes sense in a tightly trapped condensate.

\end{abstract}
\maketitle
\pacs{PACS Nos. }
\long\def\comment#1{\vskip 2mm\noindent\fbox{%
\vbox{\parindent=0cm{\small\em #1}}}\vskip 2mm}
\narrowtext
\section{Introduction}\Label{introduction}
In two previous papers on quantum kinetic theory, which we shall call QKI 
\cite{QKI} and QKII  \cite{QKII} in this paper, we focussed on 
the theory of a Bose gas in which there was
\begin{itemize}
\item[a)] No more than a small amount of Bose condensate, so that the 
interactions between the particles were not sufficiently strong to produce a 
significant modification to the excitation spectrum of the Bose gas;
\item[b)] No trapping potential, so that the system was spatially homogeneous.
\end{itemize}
Such a treatment gives insight into the the statistical aspects of quantum 
kinetics, but cannot be applied to the condensates presently being produced
\cite{JILA,MIT,RICE,AUSTIN,ROWLAND,STANFORD}, in which there is very strong 
condensation of an alkali metal vapor in a rather tight trap.

In this work the focus will be on strongly condensed trapped systems, with a 
methodology based on that of QKI, but somewhat simplified in order to exhibit 
the major features of the interaction of a Bose-condensate with a vapor of 
non-condensed particles.  The basis of our method is to divide the condensate 
spectrum into two bands; the {\em condensate band}, in which the presence of 
the condensate significantly modifies the excitation spectrum, and the
{\em non-condensate band}, whose energy levels are sufficiently high for the 
interaction with the condensate to be negligible.  

In Fig.1(a) we depict the kind of trap for which this division is very 
clear---a rather wide trap, in the center of which is a small tight trap, with 
a few energy levels.  In contrast, in Fig.1(b) we depict a harmonic trap, in 
which the division is not so obvious.  In fact, however, the three dimensional 
nature of the trap, in which the density of states increases quadratically with 
energy, also yields a relatively small number of isolated levels at low 
energies, which rapidly approaches a continuum, so the dramatic difference 
between the two traps is more apparent than real.  Current experiments all use 
harmonic traps, but the ``ideal''  trap of Fig.1(b) would in principle be 
better adapted to studies of the condensate itself.

The dynamics of the non-condensate band is thus well described by the kind of 
formalism used in QKI, with modifications to take account of the fact that we 
have also introduced a trapping potential $ V_T({\bf x})$.
On the other hand,
the dynamics of the condensate band is well described by the energy spectrum
which arises from Bogoliubov's method \cite{Bogoliubov}, as adapted to apply to 
the case of particles in a trap \cite{Fetter}.  This kind of dynamics has been 
well studied by many workers in the past two years
\cite{Fetter,Lewenstein-You,Burnett}, and the equation of motion for the 
condensate wavefunction is now generally verified both theoretically and 
experimentally to be the time-dependent Gross-Pitaveskii equation
\cite{GP refs old and new}.

The aim of this paper is to develop and use methods to describe the interaction 
between the condensate band and the non-condensate band.  The results of our 
investigations can be summarized as follows:
\begin{itemize}
\item[i)] The {\em non-condensate band\/} is taken to be fully thermalized, in 
the sense that at any point $ {\bf u},{\bf k}$ in phase space there is a local 
particle phase-space density $ F({\bf u},{\bf k})$, and the space time 
correlation functions of this phase space density have a weakly damped 
classical form.
\item[ii)] The dynamics of the condensate band, in interaction with the 
thermalized non-condensate band, is given by a master equation
(\ref{II4.11a}--\ref{II4.11g}), in which the interaction terms are of three 
kinds:
\begin{itemize}
\item[(a)] Two particles in the non-condensate band collide, and after the 
collision, one of these particle remains within the condensate band; and there 
is also of course the corresponding time-reversed process.
\item[(b)] A particle in the non-condensate band collides with a particle in 
the condensate band, and exchanges energy with it, but does not enter the 
condensate band.
\item[(c)] A particle in the non-condensate band collides with one in the 
condensate band, and after the collision both particles are left in the 
condensate band; and again there 
is also of course the corresponding time-reversed process.
\end{itemize}
\item[iii)] By using the particle-number-conserving Bogoliubov method 
\cite{trueBog} developed by one of us, the expressions which arise in the 
master equation can be re-written in a form in which the processes are 
described in terms of
\begin{itemize}
\item[(a)] Creation and destruction of particles; thus if $ N$ is the total 
number of particles in the condensate band, we introduce operators $ A^\dagger, 
A$ which change $ N$ by $ \pm1$.
\item[(b)] Creation and destruction of quasiparticles, which are to be regarded 
as quantized oscillations of the whole body of $ N$ particles in the condensate 
band.  
\end{itemize}
Thus there is a clear distinction made between creation and destruction of 
{\em particles} $ (N\to N\pm1)$, and the creation and destruction of {\em 
quasiparticles}, which are mere excitations, 
and not particles.  This distinction follows from the particle number 
conserving Bogoliubov method \cite{trueBog}, and has not been made clear in 
nearly all\footnote{A notable exception is the work of Girardeau and Arnowitt
\cite{Girardeau1}} 
earlier treatments of Bogoliubov's method, which blur this distinction, and 
give the misleading impression that the Bogoliubov excitation spectrum can 
only be obtained if strict conservation of particle number is abandoned.  We 
emphasize here that the particle number conserving Bogoliubov method gives 
exactly the same results for the excitation spectrum as the usual version; 
however, because particle number conservation is maintained, it is possible to 
specify both wavefunctions and energy eigenvalues for states with a definite 
energy and a definite number of particles.
\item[iv)] It is possible to split off the largest terms in the master 
equation, and approximate the description by a single basic master equation for 
$ N$, the number of particles in the condensate band, and thus arrive at a 
description of the growth of a condensate from the vapor.  The very simplest
version of this description, in which all fluctuations are neglected, takes 
the 
form of a single differential equation
\begin{eqnarray}\label{intro1}
\dot N = 2W^+(N)\left\{\left(1-e^{[\mu-\mu(N)]/k_{B}T}\right)N+1\right\}
\end{eqnarray}
in which $ W^+(N)$ is a certain kinetic coefficient, for which we can
give an approximate form, and $ \mu$ and $ \mu(N)$ are respectively the vapor 
chemical potential and the chemical potential of the condensate wavefunction 
for $ N $ particles, as given by solving the time-independent Gross-Pitaevskii 
equation.

\item[v)]  Detailed descriptions of the dynamics of fluctuations can be given.  
The stationary solution for the condensate density operator is a {\em mixture}
({\em not} a quantum-mechanical superposition) of states with different numbers 
$ N$ of particles.  For each such state, the wavefunction of the condensate is
$ \xi_N({\bf x})$, the solution of the the time-independent Gross-Pitaevskii
equation for $ N$ particles.
\item[iv)] A description of both phase diffusion of the condensate and of the 
decay of excitations can also be given, and these will appear in a subsequent 
paper. \cite{QKIV}.
\end{itemize}

The description given in this paper is a simplified one, in which the 
non-condensate band is treated as stationary, so that it does constitute a 
genuine ``heat-bath'', characterized by a fixed temperature $T$ and a fixed 
chemical potential $ \mu$.  This is not an essential feature of the basic 
quantum kinetic description, which, in its full form, treats the two bands as 
having their own dynamics as well as being coupled together.  The full quantum 
kinetic theory will be presented in a subsequent paper, QKV.  It turns out to 
be a rather intricate process to derive such a description, even though the 
essential result, as used in this paper, is that the non-condensate band is 
described by an equation of the Uehling-Uhlenbeck form \cite{UU}, with 
correlation functions of a thermal form.

\section{Description of the system}\Label{Desc}

\subsection{Hamiltonian and notation}
As in QKI, we consider a set of Spin-0 Bose particles, described by the
Hamiltonian
\begin{eqnarray}\Label{Desc1.1}
H=H_{\rm kin}+H_I + H_T,
\end{eqnarray}
in which
\begin{eqnarray}\Label{Desc1.2}
H_{\rm kin}=\int d^3{\bf x}\,\psi ^{\dagger }({\bf x})
\left(- {\hbar ^2\over2m}\nabla ^2\right) \psi ({\bf x})   
\end{eqnarray}
and
\begin{eqnarray}\Label{Desc1.3}
H_I&=&{1\over 2}\int d^3{\bf x}\int d^3{\bf x}'\psi ^{\dagger }({\bf x})
\psi ^{\dagger }( {\bf x}') u( {\bf x}-{\bf x}') 
\psi ( {\bf x}') \psi ( {\bf x}) . 
\nonumber \\  
\end{eqnarray}
The potential function $ u( {\bf x}-{\bf x}')$ is a c-number and is as usual 
not the true interatomic potential, but rather a short range 
potential---approximately a delta function---which reproduces the correct 
scattering length.  This enables the Born approximation to be applied, and thus 
simplifies the mathematics considerably. \cite{Born approx}

The term $ H_T$ arises from a trapping potential, and is written as 
\begin{eqnarray}\Label{IIT1}
H_T =  \int d^3 {\bf x}\,V_T({\bf x)}\psi^\dagger({\bf x})\psi({\bf x}).
\end{eqnarray}
Thus, this is the standard second quantized theory of an interacting Bose
gas of particles with mass $ m $.


\subsection{Condensate and non-condensate bands}
 When there is strong condensation in a trap, the lowest energy level of 
the trap develops a macroscopic occupation, and modifies the other energy 
levels in the trap.  If the trap is quite tight, the lower energy levels are 
reasonably well separated, as is indeed experimentally the case 
\cite{JILA,MIT,RICE}.  The 
concept of a ``tight'' trap is however not absolute; a trap may be tight for 
the lower energy levels, but quite broad for the higher energy levels.
The essential distinction is the density of states, which increases like $ E^2$
for a harmonic trap, leading to very closely spaced levels for quite moderate 
energy.
In addition, one must also note that the modification of the energy levels 
induced by the condensate affects only the lower levels significantly.

The procedure which follows from these is to divide the particle states 
into a {\em condensate band } $ R_C$ and {\em non condensate band} $ R_{NC}$, 
and to perform a corresponding resolution of the field operator in the form
$ \psi({\bf x}) = \phi({\bf x}) + \psi_{NC}({\bf x})$.  The separation between 
the two bands is in terms of energy; we choose to define the division in such 
a 
way that the energies of all particles in $ R_{NC}$ are sufficiently large that 
there is essentially no effect on them by the presence of the condensate in 
the 
lowest energy levels.  We will describe $ R_C$ in terms of discrete trap 
energy 
levels, while $ R_{NC}$ will be taken as being essentially thermalized, with 
correlation function given by the usual equilibrium formulae.

\subsubsection{Separation of condensate and non-condensate parts of the full 
Hamiltonian}
We can now write the full Hamiltonian in a form which separates the three 
components; namely, those which act within $ R_{NC}$ only, those which act 
within $ R_C $ only, and those which cause transfers of energy or population 
between $ R_C$ and $ R_{NC}$.  Thus we write
\begin{eqnarray}\Label{H1}
H =  H_{NC} + H_{0} +H_{I,C}
\end{eqnarray}
in which $ H_{NC}$ is the part of $ H$ which depends only on $ \psi_{NC}$, 
$ H_0$ is the part which depends only on $ \phi$, and that part of the 
interaction Hamiltonian which involves both condensate band and non-condensate 
band operators
is called $ H_{I,C}$, and can be written
\begin{eqnarray}\Label{HIC}
H_{I,C}\equiv H^{(1)}_{I,C}+H^{(2)}_{I,C}+H^{(3)}_{I,C},
\end{eqnarray}
where the individual terms in $ H_{I,C}$ are the terms involving operators from 
both bands, which cause 
transfer of energy and/or particles between $ R_C$ and $ R_{NC}$. We call the 
parts involving one $ \phi$ operator
\begin{eqnarray}\Label{HIC1}
&& H^{(1)}_{I,C}\equiv 
\nonumber \\ \nonumber
&&\,  \int d^3{\bf x}\int d^3{\bf x}'u({\bf x}-{\bf x}')
\psi_{NC}^\dagger({\bf x})\psi_{NC}^\dagger({\bf x}')\psi_{NC}({\bf x})
\phi({\bf x}') 
\\ \nonumber && \, + \int d^3{\bf x}\int d^3{\bf x}'u({\bf x}-{\bf x}')
\phi^\dagger({\bf x})\psi_{NC}^\dagger({\bf x}')\psi_{NC}({\bf x})
\psi_{NC}({\bf x}').
\nonumber \\
\end{eqnarray}
The parts involving  two $ \phi $ operators are called
\begin{eqnarray}\Label{HIC2}
&& H^{(2)}_{I,C}\equiv 
\nonumber \\ \nonumber
&&\,  \int d^3{\bf x}\int d^3{\bf x}'u({\bf x}-{\bf x}')
\psi_{NC}^\dagger({\bf x})\phi^\dagger({\bf x}')\psi_{NC}({\bf x})
\phi({\bf x}')
\nonumber\\
&&+ \int d^3{\bf x}\int d^3{\bf x}'u({\bf x}-{\bf x}')
\phi^\dagger({\bf x})\psi_{NC}^\dagger({\bf x}')\psi_{NC}({\bf x})
\phi({\bf x}') 
\nonumber\\
&&+{1\over 2} \int d^3{\bf x}\int d^3{\bf x}'u({\bf x}-{\bf x}')
\psi_{NC}^\dagger({\bf x})\psi_{NC}^\dagger({\bf x}')\phi({\bf x})
\phi({\bf x}')
\nonumber\\
&&+ {1\over 2}\int d^3{\bf x}\int d^3{\bf x}'u({\bf x}-{\bf x}')
\phi^\dagger({\bf x})\phi^\dagger({\bf x}')\psi_{NC}({\bf x})\psi_{NC}({\bf 
x}'). \nonumber \\
\end{eqnarray}
The parts involving  three $ \phi $ operators are called
\begin{eqnarray}\Label{HIC3}
&& H^{(3)}_{I,C}\equiv 
\nonumber\\ \nonumber
&&\,  \int d^3{\bf x}\int d^3{\bf x}'u({\bf x}-{\bf x}')
\phi^\dagger({\bf x})\phi^\dagger({\bf x}')\phi({\bf x})
\psi_{NC}({\bf x}') 
\\ \nonumber &&\, + \int d^3{\bf x}\int d^3{\bf x}'u({\bf x}-{\bf x}')
\psi_{NC}^\dagger({\bf x})\phi^\dagger({\bf x}')\phi({\bf x})
\phi({\bf x}').
\nonumber 
\\
\end{eqnarray}

\section{Derivation of the master equation}
Using the separation into condensate band and non-condensate band operators we 
can now proceed to develop a master equation.  We have developed a full quantum 
kinetic description, in which the atoms in $ R_{NC}$ are treated by a wavelet 
method as in QKI, and the atoms in $ R_C$ in terms of the wavefunctions 
corresponding to the energy levels of the trap, modified by the presence of a 
condensate in the lowest level.  This leads to a description which is very 
comprehensive, but one whose complexity would obscure the essential aspects of 
condensate dynamics which are the subject of this paper.  This description will 
be published separately.

The present description will assume that $ R_{NC}$ is fully thermalized, and 
thus the atoms in these levels will be treated simply as a heat bath and 
source of atoms for the levels in $ R_C$, which provide a description of the 
dynamics of the condensate treated as an open system.  
The simplest assumption is that $ \rho_{NC}$ has a thermodynamic equilibrium 
form given by
\begin{eqnarray}\Label{nc1}
 \rho_{NC}&=& \exp\left(\mu N_{NC}-H_{NC}\over k_{B}T\right)
\end{eqnarray}
This gives rise to 
relatively simple equations for the kinetics of the condensate under the 
restriction that the density operator for the non-condensed atoms is clamped
in this form.
A more practical thermalization requirement is that of {\em local} 
thermalization. This is defined by requiring the field operator correlation 
functions to have a thermal form locally, and to have factorization properties 
like those which pertain in equilibrium.  Time-dependence of $ \rho_{NC} $ can 
be included as well, as long as this happens on a much slower time scale than 
than those which arise from the master equation which we derive.  The precise 
nature of these local equilibrium requirements will be specified in
Sect.\ref{Correlation}

\subsection{Formal derivation of the master equation}
The derivation of the master equation follows a rather standard methodology.
We  project out the dependence on the 
non-condensate band by defining the condensate density operator as
\begin{eqnarray}\Label{proj1}
\rho_C &=& {\rm Tr}_{NC}(\rho)
\end{eqnarray}
and a projector $ {\cal P}$ by
\begin{eqnarray}\Label{proj2}
{\cal P}\rho & = &\rho_{NC}\otimes{\rm Tr}_{NC}(\rho)
\\
&=& \rho_{NC}\otimes\rho_C ,
\end{eqnarray}
and we also use the notation
\begin{eqnarray}\Label{defQ}
{\cal Q}\equiv 1-{\cal P}
\end{eqnarray}

The equation of motion for the full density operator $ \rho$ is the von Neumann 
equation
\begin{eqnarray}\Label{H8}
\dot \rho &=& -{i\over \hbar}[H_{NC} + H_0 +H_{I,C}, \rho]
\\ &\equiv &  ({\cal L}_{NC}+{\cal L}_{0}+{\cal L}_{I,C} ) \rho .
\end{eqnarray}
We use the Laplace transform notation for any 
function $ f(t)$ 
\begin{eqnarray}\Label{Laplace}
\tilde f(s) &=& \int_0^\infty e^{-st}f(t)\,dt.
\end{eqnarray}
We then use standard methods to write the master equation for 
$v(t)\equiv {\cal P}\rho(t)$ as
\begin{eqnarray}\Label{H9}
&&\quad  s \tilde v(s) -  v(0) =
{\cal L}_{0}\tilde v(s) 
+ {\cal P}{\cal L}_{I,C}\tilde v(s)
\nonumber \\ &&\qquad
+{\cal P}{\cal L}_{I,C}[s- {\cal L}_{NC}-{\cal L}_{0} -{\cal Q}{\cal 
L}_{I,C}]^{-1}
{\cal Q}{\cal L}_{I,C}\tilde v(s).
\end{eqnarray}
In this form the master equation is basically exact.  We shall make the 
approximation that the kernel of the second part, the $ [\quad]^{-1}$ term, can
be approximated by keeping only the terms which describe the basic 
Hamiltonians within $ R_C$ or $ R_{NC}$, namely the terms $ {\cal L}_{NC} $ 
and 
$ {\cal L}_{0}$.  We then invert the Laplace transform, and make a Markov 
approximation to get
\begin{eqnarray}\Label{met}
\dot v(t)& =& ({\cal L}_0+{\cal P}{\cal L}_{I,C})v(t)
\nonumber \\
&+&
\left\{{\cal P}{\cal L}_{I,C}\int_0^td\tau\,
\exp\left\{({\cal L}_0+{\cal L}_{I,C})\tau\right\}
{\cal Q}{\cal L}_{I,C} v(t)\right\}.
\nonumber \\
\end{eqnarray}
There are now a number of different terms to consider.

\subsubsection{Forward scattering terms}
These arise from the term 
${\cal P}{\cal L}_{I,C} v(t) $,  
and lead to a Hamiltonian term of the form
\begin{eqnarray}\Label{H11}
H_{\rm forward}&\equiv&
\int d^3{\bf x} \int d^3{\bf x'}\,u({\bf x}-{\bf x}')
\nonumber\\
&&\times{\rm Tr}_{NC}\bigg\{\psi^\dagger_{NC}({\bf x)}\psi_{NC}({\bf x})
\phi^\dagger({\bf x}')\phi({\bf x}')
\nonumber\\
&&+
\psi^\dagger_{NC}({\bf x)}\psi_{NC}({\bf x}')
\phi^\dagger({\bf x}')\phi({\bf x})\bigg\}
 \end{eqnarray}
The terms involving the $ \phi$ operators represent essentially two effects.  
The first is the effect of the mean density of the condensate band on the 
energy levels of the non-condensate band.  By definition of the non-condensate 
band, this effect is negligible.  They also represent an effect of the average
non-condensate density on the condensate.  This is also small, but can be 
included explicitly in our formulation of the master equation.

In  (\ref{H11}) one can see clearly the ``Hartree'' term in the second line, 
and 
the ``Fock'' term in the third line.  However in the remainder of this paper 
we shall use  the delta function form of the interaction potential
$ u({\bf x}-{\bf x}')\to u\delta({\bf x}-{\bf x}')$, and the difference 
between these terms and the second and third lines of (\ref{HIC2}) disappears.  
None of our results will depend heavily on the validity of this approximation, 
but the formulae do become considerably more compact.

\subsubsection{Interaction between $ R_C$  and $ R_{NC}$}
We now examine the term involving $ H^{(1)}_{I,C}$ as defined in
(\ref{HIC1}), which contain one
$ \phi({\bf x})$ or $ \phi^\dagger({\bf x})$; explicitly, the term in 
Hamiltonian can be written
\begin{eqnarray}\Label{bba701}
H^{(1)}_{I,C}&=&
\int d^3{\bf x}\,Z_{3}({\bf x})\phi^\dagger({\bf x})
+ {\rm h.c.}
\end{eqnarray}
In this equation we have defined a notation
\begin{eqnarray}\Label{bba802}
Z_{3}({\bf x}) 
&=& u \psi^\dagger_{NC} ({\bf x})\psi_{NC} ({\bf x})\psi_{NC} ({\bf x}).
\end{eqnarray}
Substituting into the master equation (\ref{met}), terms arise which are of the 
form
\begin{eqnarray}\Label{bba8}
&&-{1\over\hbar^2}\int d^3{\bf x}\int d^3{\bf x'}\int_0^\infty d\tau\,
{\rm Tr}\left\{Z_{3}({\bf x})Z^\dagger_{3}({\bf x}',-\tau)\rho_{NC}\right\}
\nonumber \\
&&\qquad \times \phi^\dagger({\bf x}) \phi({\bf x'},-\tau)
\rho_C(t).
\end{eqnarray}
Here we have used the notation
\begin{eqnarray}\Label{notations}
Z_3({\bf x},t)& =& e^{iH_{NC}t/\hbar}Z_3({\bf x}) e^{-iH_{NC}t/\hbar}
\\
\phi({\bf x},t)& =& e^{iH_{0}t/\hbar}\phi({\bf x}) e^{-iH_{0}t/\hbar}
\end{eqnarray}

We can now transform the term involving the $ \phi$'s into something more 
tractable, though perhaps more abstract, by expanding in eigenoperators of the 
commutator $ [H_0,\ ]$:  Thus we write
\begin{eqnarray}\Label{bba11}
\phi({\bf x}) = \sum_m X_m({\bf x})
\end{eqnarray}
where the operators $ X_m({\bf x})$ are eigenoperators of $ H_0$
\begin{eqnarray}\Label{bba10}
[H_0, X_m({\bf x})] = - \hbar \epsilon_m X_m({\bf x}).
\end{eqnarray}
Here $ \epsilon_m$ may be 
positive or negative, but $ m$ is a label which uniquely specifies the 
eigenvalue $ \epsilon_m$.  The quantities $ \epsilon_m$ give the possible 
frequencies corresponding to transitions between condensate energy
levels, and will be called the {\em transition frequencies}.
\subsubsection{The nature of the operators $ X_m$}
To compute the transition operators $ X_m$, one needs to know all of the 
eigenstates of $ H_0$.  Thus, if these eigenstates are called 
$| l, c\rangle $, with eigenvalue $ E_l$, then
\begin{eqnarray}\Label{n1}
X_m({\bf x}) &=& \sum_{ll'\atop cc'}\delta(\epsilon_m, E_l-E_{l'})
| l', c'\rangle \langle l', c'|\phi({\bf x})| l, c\rangle \langle l, c|.
\nonumber \\
\end{eqnarray}
This expression is of a formal nature, since the full calculation of the 
energy eigenstates and the matrix elements of the field operator is a  
formidable task.  However, by using the modified Bogoliubov method of 
\cite{trueBog}, this can be made into a tractable task, as will be explained in
Sect.\ref{Bog}.
\subsubsection{Random phase and rotating wave approximations}
\Label{RWA RPA}
Because the two operators $ \phi$ and $ \phi^\dagger$ both turn up, there will 
be a double summation $ \sum_{mm'}X_m^\dagger X_{m'}$.
To get a master equation it is necessary to ensure that only the $ m=m'$ terms 
exist.  This is normally achieved in quantum optical situations via the {\em 
rotating wave approximation},
which effectively eliminates the unwanted $ m\ne m'$ terms on the grounds that
they are rapidly oscillating, unlike the $ m=m'$ terms.  
Perhaps more 
compelling in the situation here is the fact that the diagonal (non-rotating) 
terms have a consistent 
sign, while the off-diagonal terms---of whiche there are very many---do not, 
and will thus tend to cancel.  This 
is effectively a {\em random phase approximation}.
Hence we will keep only the diagonal $ m=m'$ terms in what follows.
\subsubsection{\Label{Correlation}Correlation functions of $ R_{NC}$}
The terms involving $ Z_3, Z_3^\dagger$ are averaged over $ \rho_{NC}$, which 
is quantum Gaussian, but is modified by the presence of the interaction. 
A full treatment of these correlation functions requires the quantum kinetic 
theory of the non-condensate band, and this will be done in \cite{QKV}.  
However this treatment is quite intricate, and yields a result which, in the 
approximation that the non-condensate is regarded as a thermal undepleted bath,
is really quite standard and well-accepted.
This takes the form\begin{itemize}
\item[i)] We may make the replacement
\begin{eqnarray}\Label{GaussFactor}
&&\langle Z_3({\bf x}) Z^\dagger_3({\bf x}',-\tau)\rangle
\nonumber\\
&&\to  2\langle\psi_{NC}({\bf x})\psi^\dagger_{NC}({\bf x}',-\tau)\rangle
\langle\psi_{NC}({\bf x})\psi^\dagger_{NC}({\bf x}',-\tau)\rangle
\nonumber\\ 
&&\times
\langle\psi^\dagger_{NC}({\bf x})\psi_{NC}({\bf x}',-\tau)\rangle
\end{eqnarray}
(Terms involving 
$\langle\psi^\dagger_{NC}({\bf x})\psi_{NC}({\bf x}'\rangle$ etc. do arise 
in principle, but give no contribution to the final result because of energy 
conservation considerations.)

\item[ii)] The time-dependence is needed only for small $ \tau$, and in this 
case we 
can make an appropriate replacements in terms of the one-particle Wigner 
function $ F({\bf u},{\bf K})$
\begin{eqnarray}\Label{Wig1}
&&\left\langle\psi^\dagger_{NC}\left({\bf u}+{{\bf v}\over2}\right)
\psi_{NC}\left({\bf u}-{{\bf v}\over2},-\tau\right)\right\rangle
\nonumber\\
&&\,\,
\approx {1\over (2\pi)^3}\int_{R_{NC}} d^3{\bf K}\, F({\bf u},{\bf K})e^{-i{\bf 
K}\cdot{\bf v}}
e^{i\omega({\bf K},{\bf u})\tau}
\nonumber\\
\\
\label{Wig2}
&&\left\langle\psi_{NC}\left({\bf u}+{{\bf v}\over2}\right)
\psi^\dagger_{NC}\left({\bf u}-{{\bf v}\over2},-\tau\right)\right\rangle
\nonumber\\
&&\,\,
\approx {1\over (2\pi)^3}\int_{R_{NC}} d^3{\bf K}\,
\left[ F({\bf u},{\bf K})+1\right]
e^{i{\bf K}\cdot{\bf v}}
e^{-i\omega({\bf K},{\bf u})\tau}
\nonumber \\
\end{eqnarray}
where
\begin{eqnarray}\Label{omegaku}
\hbar\omega({\bf K},{\bf u})= {\hbar^2{\bf K}^2\over 2m}+V_T({\bf u}).
\end{eqnarray}
Since the range of all the $ {\bf K}$ integrals is restricted to $ R_{NC}$, 
it is implicit that in all integrals that
$\hbar\omega({\bf K},{\bf u})>E_R $.
\item[iii)] This approximation is valid in the situation where
 $ F({\bf K},{\bf u})$ is a smooth function of its arguments, and can be 
regarded as a local equilibrium assumption for particles moving in a potential
which is comparatively slowly varying in space.
\end{itemize}


\widetext
\subsubsection{Final form of the term}
Using these factorizations, the term (\ref{bba8}) becomes
\begin{eqnarray}\Label{bba9}
-\int d^3{\bf x} d^3{\bf x}'\,{\cal G}^{(-)}({\bf x},{\bf x}',\epsilon_m)
\sum_{m}X^\dagger_{m}({\bf x})X_m({\bf x}')\rho_C(t)
\end{eqnarray}
in which
\begin{eqnarray}\Label{calg}
{\cal G}^{(-)}\left({\bf u}+{{\bf v}\over2},
               {\bf u}-{{\bf v}\over2},\omega\right)
&=&{ u^2\over(2\pi)^8\hbar^2}\int d^3{\bf K}_1\int d^3{\bf K}_2\int d^3{\bf 
K}_3
\left[F({\bf K}_1,{\bf u})+1\right]\left[F({\bf K}_2,{\bf u})+1\right]
F({\bf K}_3,{\bf u})
\nonumber\\&& \times
e^{i({\bf K}_1+{\bf K}_2-{\bf K}_3)\cdot{\bf v}}
\delta^P(\omega_1+\omega_2-\omega_3-\omega)
\\ \Label{rimag}
&\equiv&
 G^{(-)}\left({\bf u}+{{\bf v}\over2},{\bf u}-{{\bf v}\over2},\omega\right)
-i  G_i^{(-)}\left({\bf u}+{{\bf v}\over2},{\bf u}-{{\bf v}\over2},\omega
\right)
\end{eqnarray}

\narrowtext
In the above we use the notation
\begin{eqnarray}\Label{bba1301}
\delta^p(x)& = & {1\over\pi}\int_0^\infty e^{-ix\tau}d\tau
\\ 	&=& \delta(x) - {i\over\pi}{{\rm P}\over x}.
\end{eqnarray}
The imaginary part in (\ref{rimag}) gives rise to level shifts, which we shall 
neglect in what follows, both for simplicity and because they are likely to be 
very small.

\subsubsection{Terms involving two $ \phi$ operators}\Label{2 phi}
Of these terms, only those involving on $ \phi$ and one 
$ \phi^\dagger$  can yield a resonant term, corresponding to scattering of a 
non-condensate particle by the condensate. The effect of the non-resonant terms 
would be a shifting of the energy levels of $ R_{NC}$, and by definition of 
the non-condensate band, this is negligible.

Thus we arrive at an approximation to $ H^{(2)}_{I,C} $ in the form of 
 a term
\begin{eqnarray}\Label{bba1302}
H^{(2)}_{I,C} &\approx&
2\int d^3{\bf x}\,Z_{2}({\bf x})\phi^\dagger({\bf x})\phi({\bf x})
+{\rm h.c.}
\end{eqnarray}
where
\begin{eqnarray}\Label{bba1303}
Z_{2}({\bf x})&=&  u \psi_{NC} ({\bf x})\psi_{NC}^\dagger ({\bf x}).
\end{eqnarray}
The term analogous to (\ref{bba8}) in the master equation becomes
\begin{eqnarray}\Label{bba14}
&& -{4\over\hbar^2}
\int_0^\infty d\tau\,\phi^\dagger({\bf x})\phi({\bf x})
\phi({\bf x}',-\tau)\phi^\dagger({\bf x}',-\tau)
\nonumber \\
\qquad
&&
\times\int d^3{\bf x}\int d^3{\bf x'}
\langle Z_{2}({\bf x})Z^\dagger_{2}({\bf x'},-\tau)\rangle
 \rho_{C}(t).
\end{eqnarray}
We can write now
\begin{eqnarray}\Label{bba1401}
\phi({\bf x})\phi^\dagger({\bf x})& =&
\sum_r\left(\sum_{m,n}X_m({\bf x})X^\dagger_n({\bf x})
\delta(\epsilon_m-\epsilon_n, \beta_r)\right)
\\
&\equiv& \sum_rU_r({\bf x}).
\end{eqnarray}
Where $ \beta_r$ represents the frequency associated with $ U_r({\bf x}) $.
Inserting these into (\ref{bba14}), we get
\begin{eqnarray}\Label{bba1402}
&&-{4\over\hbar^2}\int_0^\infty d\tau\,\int d^3{\bf x}\int d^3{\bf x'}\,
\langle Z_{2}({\bf x})Z^\dagger_{2}({\bf x'},-\tau)\rangle
\nonumber\\
&&\times\sum_r e^{i\beta_r\tau}U_r^\dagger({\bf x})U_r({\bf x'})
\rho_{C}(t).
\end{eqnarray}

\subsubsection{Terms involving three $ \phi$ operators}\Label{3 phi}
The part of the interaction Hamiltonian can be written
\begin{eqnarray}\Label{bba14010101}
H^{(3)}_{I,C} &=&
 \int d^3{\bf x}\psi_{NC}({\bf x})
\phi^\dagger({\bf x})\phi^\dagger({\bf x})\phi({\bf x})
+{\rm h.c.}
\end{eqnarray}
The master equation term this time takes the form
\begin{eqnarray}\Label{bba140102}
&& -{1\over\hbar^2}\int d^3{\bf x}\int d^3{\bf x'}\,
\psi_{NC}({\bf x})\psi_{NC}^\dagger({\bf x'})
\phi^\dagger({\bf x})\phi^\dagger({\bf x})\phi({\bf x})
\nonumber \\
&&\,\times
 \int_0^\infty d\tau\,
\phi^\dagger({\bf x}',-\tau)\phi({\bf x}',-\tau)\phi({\bf x}',-\tau)
 \rho_{C}(t).
\end{eqnarray}
We can write now
\begin{eqnarray}\Label{bba140103}
&&\phi^\dagger({\bf x})\phi({\bf x})\phi({\bf x})
\nonumber \\&&\quad  =
\sum_r\left(\sum_{m,n,p}X^\dagger_m({\bf x})X_n({\bf x})X_p({\bf x})
\delta(\epsilon_m-\epsilon_n-\epsilon_p, \beta'_r)\right)
\nonumber \\ \\
&&\quad \equiv \sum_rV_r({\bf x}).
\end{eqnarray}
Inserting these into (\ref{bba140102}), we get
\begin{eqnarray}\Label{bba140104}
&& -{1\over\hbar^2} \int_0^\infty d\tau \int d^3{\bf x}\int d^3{\bf x'}\,
\langle \psi_{NC}({\bf x})\psi_{NC}^\dagger({\bf x'},-\tau)\rangle
\nonumber\\
&&\times\sum_r e^{i\beta'_r\tau} V^\dagger_r({\bf x})V_r({\bf x'})\rho_C(t)
\end{eqnarray}
 
\widetext
\subsection{Final form of the master equation}\Label{Sect. 6.5.2}
By carrying out the procedures used to derive the term (\ref{bba9}), we can 
finally get the master equation:{\footnotesize
\begin{mathletters}
\begin{eqnarray}\Label{II4.11a}
\dot \rho _C( t)  &=&-{i\over\hbar}   \int d^3{\bf x}\,\Bigg[
\phi^{\dagger}( {\bf x})\left(-{\hbar^2\nabla^2\over2m}+V_T({\bf x})
+2u\int d^3{\bf K}\,F({\bf K},{\bf x})
 +{1\over 2}u\phi^\dagger({\bf x})\phi({\bf x})\right)
\phi( {\bf x})\, ,\,\rho_C \Bigg] 
\\ 
 \Label{II4.11b}+\int d^3{\bf x}\int d^3{\bf x}'\Bigg(
&&
\sum_mG^{(+)}({\bf u},{\bf v},\epsilon_m)
\left(2X_m({\bf x})\rho_CX^\dagger_m({\bf x'})
-\rho_CX^\dagger_m({\bf x'})X_m({\bf x})
-X^\dagger_m({\bf x'})X_m({\bf x})\rho_C\right)
\\ \Label{II4.11c}
&+&
\sum_mG^{(-)}({\bf u},{\bf v},\epsilon_m)
\left(2X^\dagger_m({\bf x})\rho_CX_m({\bf x'})
-\rho_CX_m({\bf x'})X^\dagger_m({\bf x})
-X_m({\bf x'})X^\dagger_m({\bf x})\rho_C\right)
\\ \Label{II4.11d} 
&+&\sum_rM({\bf u},{\bf v},\beta_r)
\left(2U_r({\bf x})\rho_CU^\dagger_r({\bf x'})
-\rho_CU^\dagger_r({\bf x'})U_r({\bf x})
-U^\dagger_r({\bf x'})U_r({\bf x})\rho_C\right)
\\ 	\Label{II4.11f}
&+&\sum_rE^{(+)}({\bf u},{\bf v},\beta'_r)
\left(2V_r({\bf x})\rho_CV^\dagger_r({\bf x'})
-\rho_CV^\dagger_r({\bf x'})V_r({\bf x})
-V^\dagger_r({\bf x'})V_r({\bf x})\rho_C\right)
\\ \Label{II4.11g}
&+&
\sum_rE^{(-)}({\bf u},{\bf v},\beta'_r)
\left(2V^\dagger_r({\bf x})\rho_CV_r({\bf x'})
-\rho_CV_r({\bf x'})V^\dagger_r({\bf x})
-V_r({\bf x'})V^\dagger_r({\bf x})\rho_C\right)\Bigg).
\end{eqnarray}
\end{mathletters}
In this equation we use the notation
\begin{eqnarray}\Label{uvdef}
{\bf u}&=& ({\bf x}+{\bf x}')/2
\\
{\bf v}&=& {\bf x}-{\bf x}',
\end{eqnarray}
and define the quantities $ E^{(\pm ) }$, $ M $, $ G^{(\pm ) } $ by
\begin{mathletters}
\begin{eqnarray}\Label{II4.13a}
G^{( +) }( {\bf u},{\bf v},\omega) &=&
\frac{ u^2}{(2\pi)^8\hbar ^2}
\int d^3{\bf K}_1\int d^3{\bf K}_2\int d^3{\bf K}_3
e^{i \Delta{\bf K}_{123}\cdot {\bf v}}
F({\bf K}_1,{\bf u})
F({\bf K}_2,{\bf u})[F({\bf K}_3,{\bf u})+1]
 \delta(\Delta\omega_{123}-\omega)  
\nonumber\\ \\  \Label{II4.13b}
G^{( -) }({\bf u},{\bf v},\omega) &=& 
 \frac{ u^2}{(2\pi)^8\hbar ^2}
\int d^3{\bf K}_1\int d^3{\bf K}_2\int d^3{\bf K}_3
e^{i \Delta{\bf K}_{123}\cdot {\bf v}}
[F({\bf K}_1,{\bf u})+1]
[F({\bf K}_2,{\bf u})+1]F({\bf K}_3,{\bf u})
 \delta(\Delta\omega_{123}-\omega)  
\nonumber\\ \\ \Label{II4.13c}
M({\bf u},{\bf v},\omega) &=& 
\frac{2 u^2}{(2\pi)^5\hbar ^2}
\int d^3{\bf K}_1\int d^3{\bf K}_2
e^{i({\bf K}_1-{\bf K}_2)\cdot{\bf v}}
F({\bf K}_1,{\bf u})
[F({\bf K}_2,{\bf u})+1]
 \delta(\omega_1 -\omega _2-\omega)  
\\ \Label{II4.13e}
E^{( +) }({\bf u},{\bf v},\omega) &=& 
\frac{ u^2}{2(2\pi)^2\hbar ^2}
\int d^3{\bf K}_1 e^{i{\bf K}_1\cdot{\bf v}}
F({\bf K}_1,{\bf u})
 \delta(\omega_1 -\omega)  
 \\  \Label{II4.13f}
E^{( -) }({\bf u},{\bf v},\omega) &=& 
 \frac{ u^2}{2(2\pi)^2\hbar ^2}
\int d^3{\bf K}_1
 e^{i{\bf K}_1\cdot {\bf v} }
[F({\bf K}_1,{\bf u})+1]
\delta (\omega _1-\omega) .
\end{eqnarray}
\end{mathletters}
}\narrowtext
Here we have used the notation
\begin{eqnarray}
\Label{omega1}
\omega_i &\equiv& \omega({\bf K}_i,{\bf u})
\\ \Label{deltaK}
\Delta{\bf K}_{123}&\equiv& {\bf K}_{1}+{\bf K}_{2}-{\bf K}_{3}
\\
\Label{deltaOmega}
\Delta\omega_{123} &\equiv& \omega_{1}+\omega_{2}-\omega_{3}.
\end{eqnarray}
\subsubsection{Relation between backward and forward rates}
If we consider the case that $ \rho_{NC}$ corresponds to the thermodynamic 
equilibrium form (\ref{nc1}), we can set 
\begin{eqnarray}\Label{thermeq1}
F({\bf K},{\bf u}) &\to & \bar N({\bf K},{\bf u})
\\ \Label{thermeq2}&=&
\left\{ 
\exp\left[\left(\hbar \omega({\bf K},{\bf u}) -\mu \right)/k_{B}T\right]-1
\right\}^{-1}
\end{eqnarray}
and it follows, for example, that
\begin{eqnarray}\Label{II4.15}
&& F({\bf K}_1,{\bf u})
F({\bf K}_2,{\bf u})(F({\bf K}_3,{\bf u})+1) 
e^{-(\mu-\hbar\Delta\omega_{123} /k_{B}T)}
\nonumber \\
&&\quad
=(F_{{\bf Q}_1}({\bf u})+1)(F_{{\bf Q}_2}({\bf u})+1) F_{{\bf Q}_3}({\bf u})
\end{eqnarray}
so that in this case
\begin{eqnarray}\Label{II4.16}
G^{(+) }({\bf u},{\bf v},\omega) &=&
e^{(\mu-\hbar\omega)/k_{B}T}G^{(-) }({\bf u},{\bf v},\omega) 
\\
M({\bf u},{\bf v},\omega) &=&
e^{-\hbar\omega/k_{B}T}M({\bf u},{\bf v}, -\omega) 
\\
E^{(+) }({\bf u},{\bf v},\omega) &=&
e^{(\mu-\hbar\omega)/k_{B}T}E^{(-) }({\bf u},{\bf v},\omega) 
\end{eqnarray}

\subsection{Use of the Bogoliubov approximation for the condensate}\Label{Bog}
The master equation in the form (\ref{II4.11a}--\ref{II4.11g}) gives an 
accurate  treatment of the internal dynamics of the condensate band, with an 
approximate treatment of the coupling to the non-condensate band.  However in 
order to use it in practice, it would be necessary to compute the full 
spectrum and eigenstates of the condensate, and for this it is necessary to 
have some approximate treatment of the condensate. The Bogoliubov approximation 
is the natural tool, but in its conventional form there is the difficulty that 
the Bogoliubov eigenfunctions do not have a well defined number of particles.  
In \cite{trueBog} a treatment of the condensate spectrum was given which 
combined together the Gross-Pitaevskii equation and the Bogoliubov spectrum to 
give approximate eigenfunctions with an exact number of particles.  We shall 
use this treatment in our paper, but one should bear in mind that all that is 
needed is a description of the energy levels for each number $ N$ of atoms in 
the condensate band.  In particular, the work of Girardeau and Arnowitt
\cite{Girardeau1}, based on the pair model, also presents such a 
description in the case of an untrapped condensates, which could be ganeralized 
to this situation \cite{Girardeau2}, and this would probably yield a more 
accurate description.

This treatment can be summarized as follows for the situation here, where we 
wish to use it for the condensate band, which will involve in principle some 
adaptation to take account of the fact  that $ \phi({\bf x})$
 is expanded in basis functions which belong to $ R_{C}$.  However the changes 
involved are so minor that we will not consider them in detail here.  The major 
effect is caused by the fact that the commutator of the condensate band field 
operators is not a delta function, but takes the form
\begin{eqnarray}\Label{phi commutator}
[\phi({\bf x}),\phi^\dagger({\bf x}')]&=& g({\bf x},{\bf x}')
\end{eqnarray}
where $  g({\bf x},{\bf x}') $ is the restriction of the delta function to the 
condensate band.  This is independent of the amount of condensate, since the 
boundary between the condensate and non-condensate bands is defined in terms of 
the unperturbed eigenstates.

The operator $ \phi({\bf x})$ is 
written in the form
\begin{eqnarray}\Label{tb1}
\phi({\bf x}) = a_0\xi_{N}({\bf x}) +\sum_k a_k \xi_k({\bf x})
\end{eqnarray}
where $ \xi_{N}({\bf x}) $ is the ground state wavefunction for the condensate 
in a 
situation 
where there are $ N $ particles in the condensate band.
All quantities on the right hand side are therefore implicitly functions of $ 
N$, and this expansion is thus different for each $ N$. The wavefunctions 
$ \xi_k({\bf x}) $ together with $ \xi_{N}({\bf x}) $ form a complete 
orthonormal 
set.

The essence of the method is first to find an approximation to 
$ \phi({\bf x})$ which is valid when $ \phi({\bf x})$ acts on a highly 
condensed state, in which most of the particles are in the state represented by 
the wavefunction $ \xi_{N}({\bf x}) $.  Such a highly condensed state is 
written 
$ |n_0, {\bf n}\rangle$, where $  {\bf n} \equiv \{n_k\} $ represents the 
vector of occupation numbers of the non-condensed modes within the condensate 
band.

If $ N= n_0 +\sum_k n_k$, the state can be rewritten in a form which uses 
{\it the total number of particles $ N$} rather than $ n_0$ and we will call 
this $ |N, {\bf n}\rangle$.  One now defines the operators
\begin{eqnarray}\Label{tb2}
A|N,{\bf n}\rangle&=& \sqrt{N}|N-1,{\bf n}\rangle\\
\Label{tb3}
{\cal N}  &\equiv& A^\dagger A 
\\ \Label{tb4}
 \alpha_{ k} &=& {1\over\sqrt{\cal N}}a_0^\dagger a_{ k}   	
\end{eqnarray}
If $ N $ is very large, we can write approximations valid in the space of fixed 
$ N$
\begin{eqnarray}\Label{tb5}
a_0 &\approx & A \left(1-{\sum_k n_k \over 2 N }\right )
\\ \Label{tb6}
&\approx & A.
\end{eqnarray}
The simple form (\ref{tb6}) is adequate in almost all situations, except the 
actual computation of the eigenfunctions.
The inverse of (\ref{tb4}) is also approximately given by
\begin{eqnarray}\Label{tb7}
a_k \approx {A\alpha_k\over \sqrt{N+1}}
\end{eqnarray}
and the field operator is then approximated by
\begin{eqnarray}\Label{tb8}
\phi({\bf x})
 &\approx&
 A\left( 
\xi_{N}({\bf x}) + {1\over \sqrt{ N+1}}\sum_k \xi_k({\bf x})\alpha_k
\right)
\\ \Label{tb8a}
&\equiv &
 A\left( \xi_N({\bf x}) + {1\over \sqrt{ N+1}}\chi({\bf x})\right)
.
\end{eqnarray}
(The corresponding formulae in \cite{trueBog} use $ \sqrt{N}$, rather than $ 
\sqrt{N+1}$, which is what appears most naturally in the computations.  The 
difference is of order $ 1/N$, which is better than the accuracy of the 
expansion, but the $ \sqrt{N+1}$ form gives much more elegant formulae because 
of the cancellation with other terms---hence its use here.)
Finally we can show that the {\it phonon} operators
$ \alpha_k$, and the operator $ A $ behave approximately like a set of 
independent annihilation operators; i.e.,
\begin{eqnarray}\Label{tb9}
[\alpha_k, \alpha_k^\dagger] \approx \delta_{k,k'}.
\end{eqnarray}
It is then found that if $ \xi_N({\bf x})$ represents the ground state 
wavefunction, the following are true.

\subsubsection{Gross-Pitaevskii equation}
The ground state wavefunction $ \xi_N({\bf x})$ must be a solution of the 
time-independent Gross-Pitaevskii 
equation
\begin{eqnarray}\Label{tb10}
&&-{\hbar^2\over 2m}\nabla^2\xi_N({\bf x}) 
+V_T({\bf x})\xi_N({\bf x}) + N u \bigl |\xi_N({\bf x})\bigr |^2 
\xi_N({\bf x}) 
\nonumber\\
&&\qquad\qquad =\mu_N \xi_N({\bf x}).
\end{eqnarray} 
In fact, because of the modified commutator (\ref{phi commutator}), the precise 
equation is given by the projection of the right hand side of (\ref{tb10}) onto 
the condensate band being equal to the left hand side, but there will be very 
little practical difference introduced by this modification.  It is, roughly 
speaking, equivalent to solving the partial differential equation on a grid of 
spacing similar to the lowest wavelength in the non-condensate band.  This is 
in fact what is done in any practical procedure---the only modification is that 
we now specify that the grid not be finer than required by the restriction to 
the condensate band.

\subsubsection{Approximate Hamiltonian}
The procedure can be put in the form of an expansion in inverse powers of 
$ N^{1/2}$ by formalizing the requirement that the number of particles $ N$ be 
large and the interaction strength $ u$ be small.  To do this one sets
\begin{eqnarray}\Label{expansion}
u = \tilde u /N
\end{eqnarray}
and then develops the approximation procedure as an asymptotic expansion at 
fixed $ \tilde u$ in inverse powers of $ N^{1/2}$.  Using this procedure
the condensate band Hamiltonian can be then approximated by
\begin{eqnarray}\Label{tb11}
H_0 = N{\cal H}_1+{\cal H}_3
\end{eqnarray}
where

\widetext
\begin{eqnarray}\Label{tb12}
{\cal H}_1&=& 
-{\hbar^2\over 2m}\int d^3{\bf x}\,\xi_{N}^*({\bf x})\nabla^2\xi_{N}({\bf x})
+\int d^3{\bf x}\,\xi_{N}^*({\bf x})V_T({\bf x})\xi_{N}({\bf x})
+{\tilde u\over 2}\int d^3{\bf x}\,
\big |\xi_{N}({\bf x})\big |^4,
\\
&\equiv & E_0(N)/N
\end{eqnarray}
which is a c-number, and
{\footnotesize
\begin{eqnarray}
\Label{t5c}
{\cal H}_3 &=&
-{\hbar^2\over 2m}\int d^3{\bf x}\,\chi^\dagger({\bf x})\nabla^2\chi({\bf x})
+\int d^3{\bf x}\,\chi^\dagger({\bf x})V({\bf x})\chi({\bf x})
\nonumber \\&&
+ \int d^3{\bf x}\bigg\{
{\tilde u\over 2}\big(\xi_N({\bf x})\chi^\dagger({\bf x})\big)^2 +
{\tilde u\over 2}\big(\xi_N^*({\bf x})\chi({\bf x})\big)^2
 +
\chi^\dagger({\bf x})\chi({\bf x})\Big(
{2\tilde u}\big |\xi_N({\bf x})\big |^2- \mu_N\Big)\bigg\}
-{\tilde u\over 2}\int d^3{\bf y}\big|\xi_N({\bf y})\big|^4 .
\end{eqnarray}
}
\narrowtext
\subsubsection{Chemical potential}
The condensate chemical potential $ \mu(N)$ is determined by the requirement 
that
\begin{eqnarray}\Label{tb13}
\int d^3{\bf x}\,|\xi_{N}({\bf x})|^2 =1
\end{eqnarray}
and this will give a different $ \mu(N)$ for each $ N$: It can 
be shown that
\begin{eqnarray}\Label{tb14}
\mu(N) \approx E_0(N+1)-E_0(N).
\end{eqnarray}

\subsubsection{Bogoliubov transformation}
The Hamiltonian $ {\cal H}_3$ can be diagonalized by a Bogoliubov
transformation of the form
\begin{eqnarray}\Label{tb15}
\alpha_k = \sum_mc_{km}b_m + \sum_ms_{km}b^\dagger_m
\end{eqnarray}
and here $ b_m$ is a quasiparticle destruction operator.  We can then 
write
\begin{eqnarray}\Label{tb16}
\chi({\bf x}) = 
\sum_m\left(p_m^*({\bf x})b_m + q_m^*({\bf x})b^\dagger_m\right)
\end{eqnarray}
with
\begin{eqnarray}\Label{tb17}
p_m({\bf x}) &=& \sum_k c_{km}^*\xi_k({\bf x})
\\ \Label{ tb18}
q_m({\bf x}) &=& \sum_k s_{km}^*\xi_k({\bf x}).
\end{eqnarray}
The diagonalized Hamiltonian is then written
\begin{eqnarray}\Label{tb19}
{\cal H}_3
 = \hbar\omega_g(N) + \sum_m\hbar\epsilon_m(N) b^\dagger_mb_m.
\end{eqnarray}
The practical task of determining the values of the necessary quantities is not 
trivial, but it is at least perfectly well defined.  It does not matter, in 
principle, what functions $ \xi_k({\bf x})$ are chosen as long as they are a
complete orthonormal set in the space orthogonal to $ \xi_{N}({\bf x})$.  
However,
a choice of appropriate $ \xi_k({\bf x})$ may make the practical problem of 
diagonalizing $ {\cal H}_3$ somewhat more straightforward.  There have been a 
number of papers \cite{Lewenstein,Burnett,Javanainen} in  which numerical 
methods of diagonalizing $ {\cal H}_3 $  have been presented

\subsubsection{Relationship between ground states for $ N$ and $ N+1$ 
particles.}
The operator $ A$ does depend on $ N$ because $ a_0$ is defined by 
(\ref{tb2}) to be 
\begin{eqnarray}\Label{cor1}
a_0(N) = \int d^3{\bf x}\,\xi_{N}^*({\bf x},N)\phi({\bf x}),
\end{eqnarray}
where the explicit dependence on $ N$ of both $ \xi_{N}$ and $ a_0$ has now 
been written. 

This means that $ a_0^\dagger(N)|N, {\bf n} = {\bf 0}\rangle$
is a state with $ N+1$ particles in the wavefunction corresponding to ground 
state of the
$ N$ particle state; it is not the ground state for $ N+1$ particles.
We will introduce the notation $ |M\rangle_N$ to mean a state with $ M$ atoms 
in the wavefunction corresponding to the $ N$ particle ground state.

It can be shown that the operator $ B^\dagger(N)$ which connects the 
$ N$ and $ N+1$ ground states through
\begin{eqnarray}\Label{cor901}
B^\dagger(N)|N\rangle_N =\sqrt{N+1} |N+1\rangle_{N+1} 
\\
      B(N-1)|N\rangle_N =\sqrt{N}   |N-1\rangle_{N-1} 
\end{eqnarray}
is given  approximately by (to order $ 1/N$)
\begin{eqnarray}\Label{cor10}
B^\dagger(N) &\approx &
\left(1+{1\over\sqrt{N}}\sum_kr^*_k\alpha^\dagger_k\right)a_0^\dagger(N).
\end{eqnarray}
where
\begin{eqnarray}\Label{cor1001}
r_k = N\int d^3{\bf x}\,
                 {\partial\xi_N^*({\bf x}) \over\partial N }\xi_k({\bf x},N).
\end{eqnarray}
This means that the field operator expansion (\ref{tb8a}) now takes the form
\begin{eqnarray}\Label{cor11}
\phi({\bf x}) &\approx&
B(N)\Big(\xi_N({\bf x}) 
\nonumber\\ &&
+{1\over\sqrt{N+1}}\Big\{\chi({\bf x}) - \xi_N({\bf x})\sum_k r_k\alpha_k
\Big\}
\Big). \nonumber \\
\end{eqnarray}
It is particularly interesting to see that the corrections to the 
quasiparticle term are of the same order of magnitude as the original terms;
the correction is thus quite significant.

\subsection{Transformation of the master equation}
The Bogoliubov approximation for $ \phi({\bf x})$ (\ref{tb8a}), together with 
the expression of $ \chi({\bf x})$ as in (\ref{tb16}) has the advantage that 
it 
gives, almost directly, the operators $ X_m({\bf x})$ which were implicitly 
defined in (\ref{bba11}).  

\subsubsection{Transition operators in the master equation}
We can see that there are three kinds of transition operators as defined in 
(\ref{bba11}), but their form depends on the number of condensate band 
particles $ N$ in the state on which they act.  Thus, we can write the action 
of $ \phi$ on an arbitrary state in the $ N$ basis as
\begin{eqnarray}\Label{rc1}
\phi({\bf x}) |A\rangle_N &=& \left(X_0({\bf x})+
\sum_m\left\{X^-_m({\bf x})+X^+_m({\bf x})\right\}\right) |A\rangle_N
\nonumber \\
\end{eqnarray}
in which we have the correspondence,
with eigenvalues and eigenfunctions, as follows:
\widetext
\begin{eqnarray}\Label{f1}\nonumber
\mbox{\bf Operator} &\mbox{\bf Representation} &\mbox{\bf Eigenvalue}\\
 X_0({\bf x})  \qquad&  B(N-1)\xi_{N-1}({\bf x}) \qquad&  -\mu({N-1})  \\
 X^-_m({\bf x})  \qquad& {B(N-1) b_mf_m(N-1,{\bf x}) /\sqrt{N}} 
   	\qquad&  -\hbar\epsilon_m(N-1)-\mu(N-1)  \\
 X^+_m({\bf x})  \qquad& { B(N-1) b^\dagger_mg_m(N-1,{\bf x}) /\sqrt{N}}
   	\qquad&  \hbar\epsilon_m(N-1)-\mu({N-1})
\\
 X_0({\bf x})^\dagger  \qquad&  B(N)^\dagger\xi^*_{N}({\bf x}) \qquad&  
\mu({N})   \\
 X^-_m({\bf x})^\dagger  \qquad& { B(N)^\dagger b_m^\dagger f^*_m(N,{\bf x}) 
/\sqrt{N+1}}
   	\qquad&  \hbar\epsilon_m(N) +\mu({N})   \\
 X^+_m({\bf x})^\dagger  \qquad& {B(N)^\dagger b_mg^*_m(N,{\bf x}) /
\sqrt{N+1}} 
   	\qquad&  -\hbar\epsilon_m(N)+\mu({N})
\end{eqnarray}
\narrowtext
in which 
\begin{eqnarray}\Label{cor11a}
f_m(N,{\bf x}) &=& p_m(N,{\bf x}) -\xi_{N}({\bf x})\sum_kr_kc_{km}
\\ &=& p_m(N,{\bf x})
\\&& -N\xi_{N}({\bf x})\int d^3{\bf x}'\,p_m(N,{\bf x}')
{\partial\xi_N({\bf x}') \over\partial N },
\\ \Label{cor12}
g_m(N,{\bf x}) &=& q_m(N,{\bf x}) -\xi_{N}({\bf x})\sum_kr_ks_{km}
\\ &=& q_m(N,{\bf x})
\\&& - N\xi_{N}({\bf x})\int d^3{\bf x}'\,q_m(N,{\bf x}')
{\partial\xi_N({\bf x}') \over\partial N },
\end{eqnarray}
The importance of the corrections can now be seen---they are of the same order 
of magnitude as the direct quasiparticle terms, and represent the generation 
of 
quasiparticles as a consequence of the changing shape of the condensate.

As given above, the choice of how to write the operator $ \phi({\bf x})$ 
depends on the value of $ N$, and in a master equation we have to consider the 
action of several $ \phi$ operators on the left or right of a density operator, 
which may contain off-diagonal terms.
To see what is involved, suppose we take a term like 
$ \phi^\dagger\rho\phi$, and consider a term in $ \rho $ like
$ |N_1\rangle_{N_1}\langle N_2 |_{N_2}$.
It is soon clear that the use of the $ N_1$ 
expansions on the right and the $ N_2$ expansion on the left leads to an 
evolution operator which is extremely complicated.  However, in practice we 
will not be interested
in density operators which are far from diagonal.  This means that we will 
instead write off-diagonal
terms in the form $ |N-\nu\rangle_{N}\langle N+\nu |_{N}$, and for such terms 
the expansion appropriate to $ N $ will be the correct one.  An off diagonal 
term of this kind is an eigenfunction of the two-sided operator $ \tilde N $, 
defined by
\begin{eqnarray}\Label{tilden}
\tilde N\rho&\equiv& {1\over 2}[ \hat N,\rho]_{+}
\\ &=&    {1\over 2}[B^\dagger(N-1)B(N-1),\rho]_{+}
\end{eqnarray}
An arbitrary density operator can then be expanded in the form
\begin{eqnarray}\Label{dens1}
\rho_N = \sum_{\nu}r_{\nu} |N-\nu\rangle_N\langle N+\nu |_N.
\end{eqnarray}
We now treat the problem of expanding master equation terms of the 
kind
\begin{eqnarray}\Label{dens2}
\mbox{\rm Term 1 }&=& 2\phi({\bf x})\rho_N \phi^\dagger({\bf x}')
- [\rho_N, \phi^\dagger({\bf x'})\phi({\bf x})]_+
,
\\ 
\mbox{\rm Term 2 }&=& 2\phi^\dagger({\bf x})\rho_N \phi({\bf x}')
- [\rho_N ,\phi r({\bf x'})\phi^\dagger({\bf x})]_+
.
\end{eqnarray}
We use the the expansion of the field operator as follows; we work only to 
lowest order, that is we omit all quasiparticle terms, for which the working is 
analogous:
\begin{eqnarray}\Label{dens3}
&&\phi^\dagger({\bf x'})\phi({\bf x})\rho_N \to 
\phi^\dagger({\bf x'}) \xi_{N-1}({\bf x})B(N-1)\rho_N
\nonumber \\ 
&&\qquad
\to  \xi^*_{N-1}({\bf x'})B^\dagger(N-1)\xi_{N-1}({\bf x})B(N-1)\rho_N.
\end{eqnarray}
The use of the $ B^\dagger(N-1)$ follows because $ B(N-1)$ takes a term like
$ |N-\nu\rangle_N \to |N-1-\nu\rangle_{N-1}$, and then $ B^\dagger(N-1)$ takes 
it back to $ |N-\nu\rangle_N $
\widetext 
Using this methodology, we then find that
\begin{eqnarray}\Label{dens4}
\mbox{\rm Term 1 }&=&
 \xi_{N-1}({\bf x})\xi^*_{N-1}({\bf x}')
\left\{2B(N-1)\rho_N B^\dagger(N-1)- [\rho_N,B^\dagger(N-1)B(N-1) ]_+\right\}
,
\\ 
\mbox{\rm Term 2 } &=&
 \xi_{N}({\bf x})\xi^*_{N}({\bf x}')
\left\{2B(N)^\dagger\rho_N B(N)- [\rho_N,B(N)B^\dagger(N) ]_+\right\}
.\end{eqnarray}
In order to get a form in which the resolution of the density operator into 
terms $ \rho_N $ is not necessary we use the operator $ \tilde N$ to rewrite 
(\ref{dens4}) in a form valid for any $ \rho$
\begin{eqnarray}\Label{dens5}
\mbox{\rm Term 1 }&=&
\left\{2B
\left\{\xi_{\tilde N-1}({\bf x})\xi^*_{\tilde N-1}({\bf x}')
\rho\right\} B^\dagger- 
\left[\left\{\xi_{\tilde N-1}({\bf x})\xi^*_{\tilde N-1}({\bf x}')
\rho\right\},B^\dagger B \right]_+\right\}
,
\\ 
\mbox{\rm Term 2 }&=&
\left\{2B^\dagger
\left\{\xi_{\tilde N}({\bf x})\xi^*_{\tilde N}({\bf x}')\rho\right\}
B- 
\left[\left\{\xi_{\tilde N}({\bf x})\xi^*_{\tilde N}({\bf x}')\rho\right\},
BB^\dagger \right]_{+}\right\}
\end{eqnarray}
\narrowtext
In this form it must be understood that 
\begin{itemize}
\item $ \rho$ is expanded as a sum of terms $ \rho_N$.
\item The operators $ B$, $ B^\dagger$ are defined by
\begin{eqnarray}\Label{dens6}
B|A\rangle_N   &\equiv& B(N-1)|A\rangle_N 
\\
\langle A|_N B &\equiv& \langle A|_N B(N)
\\
B^\dagger|A\rangle_N   &\equiv& B^\dagger(N)|A\rangle_N 
\\
\langle A|_N B^\dagger &\equiv& \langle A|_N B^\dagger(N-1)
\end{eqnarray}
\end{itemize}
The terms involving quasiparticles can be similarly treated.

\subsubsection{Other transition operators}
The operators $ U_r({\bf x})$ and $ V_r({\bf x})$ are similarly able to 
expressed using these expressions.

\subsubsection{Master equation in terms of the Bogoliubov states}
We  can now write out the master equation in terms of 
six transition 
probabilities defined in terms of functions $ R^\pm$ as
\begin{eqnarray}\Label{new3}
W^{+}(N) &=& R^+(\xi_{ N},\mu_N/\hbar)
\\ \Label{new3a}
W^{-}(N+1) &=& R^-(\xi_{N},\mu_{N}/\hbar)
\\ \Label{new4}
W_m^{++}(N) &=& R^+\big(f_m,(\epsilon^m_N+\mu_N)/\hbar\big)
\\ \Label{new5a}
W_m^{--}(N+1) &=& R^-\big(f_m,(\epsilon^m_{N}+\mu_{N})/\hbar\big)
\\ \Label{new5}
W_m^{+-}(N) &=& R^+\big(g_m,(-\epsilon^m_N+\mu_N)/\hbar\big)
\\ \Label{new4a}
W_m^{-+}(N+1) &=& R^-\big(g_m,(-\epsilon^m_{N}+\mu_{N})/\hbar\big)
\end{eqnarray}
The functions
$ R^{\pm}(y,\omega)$ are defined by
{\footnotesize
\widetext
\begin{eqnarray}\Label{new1}
 R^{+}(y,\omega)&=&{u^2\over(2\pi)^5\hbar^2}\int d^3{\bf x}
\int d^3{\bf K}_1 d^3{\bf K}_2 d^3{\bf K}_3 d^3{\bf k}\,
\delta(\Delta\omega_{123}({\bf x})-\omega)
\delta({\bf K}_1 + {\bf K}_2 - {\bf K}_3 - {\bf k})
F_1F_2 (1+F_3) {\cal W}_y({\bf x},{\bf k})
\\ \Label{new2}
 R^{-}(y,\omega)&=&{u^2\over(2\pi)^5\hbar^2}\int d^3{\bf x}
\int d^3{\bf K}_1 d^3{\bf K}_2 d^3{\bf K}_3 d^3{\bf k}\,
\delta(\Delta\omega_{123}({\bf x})-\omega)
\delta({\bf K}_1 + {\bf K}_2 - {\bf K}_3 - {\bf k})
( 1+F_1)( 1+F_2)F_3 {\cal W}_y({\bf x},{\bf k})
\nonumber \\
\end{eqnarray}
}
Here we use the notation 
\begin{eqnarray}\Label{f10}
 {\cal W}_y({\bf x},{\bf k}) = {1\over (2\pi)^3}\int d^3{\bf v}\,
y^*\left({\bf x}+{{\bf v}\over 2}\right)y\left({\bf x}-{{\bf v}\over 2}\right)
e^{i{\bf k}\cdot{\bf v}}
\end{eqnarray}
to represent the Wigner function 
corresponding to the wavefunction $ y({\bf x})$.

We now use the explicit 
expression (\ref{II4.13a}) for $ G^{(+)}({\bf u},{\bf v},\omega)$, and 
substitute 
for the $ X $ operators to get
\begin{eqnarray}\Label{f9}
\ref{II4.11b}&&\to  
2B\left\{W^{-}(\tilde N)\rho\right\} B^\dagger
-\left[B^\dagger B,\left\{W^{-}(\tilde N)\rho\right\} \right]_+ 
\nonumber \\ && +
2Bb_m\left\{W_m^{--}(\tilde N)\rho\over\tilde N\right\}  b_m^\dagger B^\dagger
-\left[b_m^\dagger B^\dagger Bb_m,\left\{W_m^{--}(\tilde N)\rho\over\tilde N
\right\}
\right]_+ 
\nonumber \\ && +
2Bb^\dagger_m\left\{W_m^{-+}(\tilde N)\rho\over\tilde N\right\} b_m B^\dagger
-\left[b_m B^\dagger Bb^\dagger_m,\left\{W_m^{-+}(\tilde N)\rho\over\tilde N
\right\}
\right]_+ .
\end{eqnarray}

The reversed process leads to the term given by
\begin{eqnarray}\Label{f10a}
\ref{II4.11c}&&\to  
2B^\dagger\left\{W^{+}(\tilde N)\rho\right\} B
-\left[B B^\dagger,\left\{W^{+}(\tilde N)\rho\right\} \right]_+ 
\nonumber \\ && +
2B^\dagger b_m\left\{W_m^{+-}(\tilde N)\rho\over \tilde N+1\right\} 
 b_m^\dagger B
-\left[b_m^\dagger B B^\dagger b_m,\left\{W_m^{+-}(\tilde N)\rho\over \tilde 
N+1
\right\}\right]_+ 
\nonumber \\ && +
2B^\dagger b^\dagger_m\left\{W_m^{++}(\tilde N)\rho\over\tilde N+1\right\} b_m 
B
-\left[b_m B B^\dagger b^\dagger_m,\left\{W_m^{++}(\tilde N)\rho\over\tilde N+1
\right\}
\right]_+ .\end{eqnarray}
\narrowtext
 \subsubsection{More complex terms}
The terms involving two and three condensate field operators would become very 
complicated unless an abbreviated notation is devised.  We introduce a 
notation 
\begin{eqnarray}\Label{comp1}
B&=& S_0
 \\
 Bb_m &=& S_m
\\ 
Bb^\dagger_m &=& S_{-m}
\end{eqnarray}
so that while  $m$ is an index whose range is conventionally the positive 
integers, there is now an index $ I =0, \pm1, \pm 2,\dots $ which enumerates 
the full range of operators.  

We also introduce the notations
\begin{eqnarray}\Label{comp2}
 L_0({\bf x},N) &=&\xi({\bf x},N)
 \\\Label{comp3}
 L_m({\bf x},N) &=&f_m({\bf x},N)/\sqrt{N+1} 
\\ \Label{comp4}
 L_{-m}({\bf x},N) &=&g_m({\bf x},N)/\sqrt{N+1}
\\ \Label{comp5}
  {\cal W}_{I}(N,{\bf x},{\bf k})&=&{\cal W}_{L_I}\left({\bf x}\right) 
\end{eqnarray}
and for the frequencies
\begin{eqnarray}\Label{comp6}
\mu(N)/\hbar&=& \Omega_0(N)
\\ \Label{comp7}
\mu(N)/\hbar +\epsilon_m(N)  &=& \Omega_m(N)
\\ \Label{comp8}
\mu(N)/\hbar -\epsilon_m(N)  &=& \Omega_{-m}(N)
\end{eqnarray}
In terms of these abbreviations, we can write all the two $ \phi$ terms in 
terms of a set of rate functions
\widetext
\begin{eqnarray}\Label{rate2}
R_{IJ}(N) &=&{4u^2\over(2\pi)^5\hbar^2}\int d^3{\bf x}
\int d^3{\bf K}_1\int d^3{\bf K}_2\int d^3{\bf k}\int d^3{\bf k}'
\delta({\bf K}_1 - {\bf K}_2 - {\bf k} + {\bf k}')
\\ \nonumber
&&\times
F({\bf K}_1,{\bf x})\bigl( 1+F({\bf K}_2,{\bf x})\bigr)
{\cal W}_I(N,{\bf x},{\bf k}){\cal W}_J(N,{\bf x},{\bf k}')
\delta(\Delta\omega_{12}({\bf x})-\Omega_I(N)+\Omega_J(N))
\end{eqnarray}
and this leads to the transformation
\begin{eqnarray}\Label{comp9}
\ref{II4.11d}&\to  &
\sum_{IJ}
\left\{
2S_IS_J^\dagger \left\{R_{IJ}(\tilde N)\rho\right\} S_JS_I^\dagger 
-\left[ S_JS_I^\dagger S_IS_J^\dagger,\left\{R_{IJ}(\tilde N)\rho\right\}
\right]_+ 
\right\}
\end{eqnarray}

The three $ \phi  $ terms require forward and backward rate 
functions
{\footnotesize
\begin{eqnarray}\Label{rate3f}
R^+_{IJK}(N) &=&
{u^2\over(2\pi)^5\hbar^2}\int d^3{\bf x}
\int d^3{\bf K}_1\int d^3{\bf k}\int d^3{\bf k}'\int d^3{\bf k}''
\delta({\bf K}_1 + {\bf k} - {\bf k}'-{\bf k}'')
\nonumber\\ 
&&\times
F({\bf K}_1,{\bf x})
{\cal W}_I(N,{\bf x},{\bf k}){\cal W}_J(N,{\bf x},{\bf k}'){\cal W}_K(N,{\bf 
u},{\bf k}'')
\delta\left(\omega_{1}({\bf x})+\Omega_I(N)-\Omega_J(N)-\Omega_K(N)\right)
\\
\Label{rate3b}
R^-_{IJK}(N+1)\nonumber 
&=&
{u^2\over(2\pi)^5\hbar^2}\int d^3{\bf x}\
\int d^3{\bf K}_1\int d^3{\bf k}\int d^3{\bf k}'\int d^3{\bf k}''
\delta({\bf K}_1 + {\bf k} - {\bf k}'-{\bf k}'')
\\ 
&&\times
\bigl(1+F({\bf K}_1,{\bf x})\bigr)
{\cal W}_I(N,{\bf x},{\bf k}){\cal W}_J(N,{\bf x},{\bf k}'){\cal W}_K(N,{\bf 
u},{\bf k}'')
\delta\left(\omega_{1}({\bf x})+\Omega_I(N)-\Omega_J(N)-\Omega_K(N)\right)
\end{eqnarray}
}
and then take the form
\begin{eqnarray}\Label{comp10}
\ref{II4.11f}&\to  &
\sum_{IJK}\left\{
2S_IS_J^\dagger S_K^\dagger\left\{R^+_{IJK}(\tilde N)\rho\right\} S_KS_J
S_I^\dagger 
-\left[S_K S_JS_I^\dagger   S_IS_J^\dagger S_K^\dagger
,\left\{R^+_{IJK}(\tilde N)\rho\right\}\right]_+
\right\}
\end{eqnarray}
The reversed three $ \phi$ terms give  
\begin{eqnarray}\Label{comp10a}
\ref{II4.11g}&\to  &
\sum_{IJK}\left\{
2S_I^\dagger S_J S_K\left\{R^-_{IJK}(\tilde N)\rho\right\} S_K^\dagger S_J^
\dagger S_I 
-\left[S_K^\dagger S_J^\dagger S_I   S_I^\dagger S_J S_K,
\left\{R^-_{IJK}(\tilde N)\rho\right\}\right]_+
\right\}
\end{eqnarray}
\narrowtext


\section{Practical application  of the master equation}
The master equation using all the transition operators is a rather complex
object, and it is wise to attempt a simplification which would give only the
most significant contributions in order to get some insight into the basic 
structure predicted.  The most obvious simplification is to take into account 
only the terms which are most significant when there is a large amount of 
condensate. This means that we will consider only the terms which are of 
highest order in $ N$, and these are easily identifiable as:
\begin{itemize}
\item[a)]  Those terms arising from the first lines of (\ref{f9},\ref{f10a});
\item[b)] The terms involving only $ R_{00}(N)$ in (\ref{comp9});
\item[c)] The terms involving only $ R^{\pm}_{000}(N)$ are certainly almost 
zero because they would involve the absorption of an atom from the 
non-condensate band into the condensate level, which cannot conserve energy.  
Thus we can probably ignore these terms completely.
\end{itemize} 

\subsection{A master equation for the condensate mode alone}
For a first investigation we will consider only the terms from the first lines 
of (\ref{f9},\ref{f10a}), since the $ R_{00}(N)$ terms cannot change the 
number of particles in the condensate.
This leaves a simplified 
master equation for the condensate mode alone in the form
\begin{eqnarray}\Label{simp1}
\dot \rho &=& -{i\over \hbar}\left[H'_0,\rho\right] 
\nonumber \\
&&
+2B^\dagger\left\{W^{+}(\tilde N)\rho\right\} B
-\left[B B^\dagger,\left\{W^{+}(\tilde N)\rho\right\} \right]_+ 
\nonumber \\ &&+
2B\left\{W^{-}(\tilde N)\rho\right\} B^\dagger
-\left[B^\dagger B,\left\{W^{-}(\tilde N)\rho\right\} \right]_+ 
\end{eqnarray}
Here $ H_0' = H_0 +2u\int d^3{\bf K}\int d^3{\bf x}F({\bf K},{\bf x})
\phi({\bf x})^\dagger\phi({\bf x})$ includes the mean-field correction to the 
trapping potential, which is probably very small.

The ground state 
wavefunction is in practice very localized compared to the size of the 
cloud of atoms in the non-condensate band. This means that 
$ F({\bf K}, {\bf x})$ is very broad in 
$ {\bf x}$ compared to $ W(\xi, {\bf x}, {\bf k})$, which is comparatively 
sharply peaked at $ {\bf x}= 0$,  so that for example in the integral in 
(\ref{f9}) over $ {\bf x}$ we can simply use  $ F({\bf K}, {\bf 0})$ and 
$ \Delta\omega_{123}({\bf 0})$.  
With these approximations, the transition rate functions (\ref{new1},
\ref{new2}) are
\widetext
\begin{eqnarray}\Label{simp2}
W^+(N)&=& {u^2\over(2\pi)^5\hbar^2}
\int d^3{\bf K}_1\int d^3{\bf K}_2\int d^3{\bf K}_3\int d^3{\bf k}\,
\delta({\bf K}_1 + {\bf K}_2 - {\bf K}_3 - {\bf k})
\nonumber \\ &&\times
\delta(\Delta\omega_{123}({\bf 0})-\mu(N)/\hbar)
F({\bf K}_1,{\bf 0})F({\bf K}_2,{\bf 0})\bigl( 1+F({\bf K}_3,{\bf 0})\bigr)
|\tilde \xi_N({\bf k})|^2
\\ \Label{simp3}
W^-(N+1)&=& {u^2\over(2\pi)^5\hbar^2}
\int d^3{\bf K}_1\int d^3{\bf K}_2\int d^3{\bf K}_3\int d^3{\bf k}\,
\delta({\bf K}_1 + {\bf K}_2 - {\bf K}_3 - {\bf k})
\nonumber \\ &&\times
\delta(\Delta\omega_{123}({\bf 0})-\mu(N)/\hbar)
\bigl( 1+F({\bf K}_1,{\bf 0})\bigr)\bigl( 1+F({\bf K}_2,{\bf 0})\bigr)
F({\bf K}_3,{\bf 0})
|\tilde \xi_N({\bf k})|^2
\end{eqnarray}
\narrowtext
\noindent
in which $ \tilde \xi_N({\bf k})$ is the momentum space condensate wavefunction
\begin{eqnarray}\Label{simp4}
\tilde \xi_N({\bf k}) &=& {1\over(2\pi)^{3/2}}\int d^3{\bf x}\,
e^{i{\bf k}\cdot{\bf x}}\xi_N({\bf x})
\end{eqnarray}
\subsection{Thermal bath of non-condensate atoms}\Label{thermal}
In this case we assume that we can write
\begin{eqnarray}\Label{simp5}
F({\bf K},{\bf x}) 
\approx\left[ e^{(\hbar\omega({\bf K},{\bf x}) -\mu)/k_{B}T}-1 \right]^{-1}
\end{eqnarray}
from which one easily obtains (choosing $ V_T({\bf 0})=0$)
\begin{eqnarray}\Label{simp6}
W^+(N) =e^{\{\mu -\mu(N)\}/k_{B}T}W^-(N+1).
\end{eqnarray}
\subsubsection{Maxwell-Boltzmann approximation}
The integrals can be approximately evaluated by making a Maxwell-Boltzmann 
approximation  
$F({\bf K},{\bf x})\approx e^{-(\hbar\omega({\bf K},{\bf x}) -\mu)/k_{B}T} $.
Although this may be in some cases a drastic approximation, it has in its 
favor:
\begin{itemize}
\item[(i):] It should always give the correct order of magnitude;
\item[(ii):] The distribution is needed only for energies in the non-condensate 
band, and will not become infinite there as long as $ E_R > \mu$;
\item[(iii):] It will certainly be valid for sufficiently large $ {\bf K}$.
\end{itemize}

\noindent
Using $ u = 4\pi a\hbar^2/m$, where $ a$ is the 
$s$-wave scattering length, we get
\begin{eqnarray}\Label{prac1}
W^+(N) &=& {4m (ak_{B}T)^2 \over \pi \hbar^3}e^{2\mu/k_{B}T}
\left\{{\mu_N\over k_{B}T}K_1\left({\mu_N\over k_{B}T}\right)\right\}.
\nonumber \\
\end{eqnarray}
Here $ K_1(z)$ is a modified Bessel function---the full details of the 
approximations involved in the evaluation of $ W^{+}(N)$ are given in 
Appendix \ref{evaluations}.
\subsubsection{A fuller quantum treatment}
Evaluation of $ W^+(N)$ beyond the Maxwell-Boltzmann approximation involves a 
more serious consideration of the cut at $ E_R$ between the condensate and 
non-condensate bands.  When one makes the Maxwell-Boltzmann approximation the 
dependence on the ``cut'' at $ E_R$ is not very strong; however, when the 
full Bose-Einstein form is used this dependence becomes more severe because of 
the Bose enhancement of the distribution function at lower energies.  The 
correct treatment of this problem must take into account the particular 
situation being treated.  For example, if we are considering the growth of the 
condensate, the condensate and non-condensate bands will not be in equilibrium 
with each other, and the dependence on $ E_R$ obtained may be very misleading 
since it arises from the lower energies at which one would expect some 
modification of the pure Bose-Einstein distribution.  One is led to the 
conclusion that this issue can only be resolved by taking into account more 
details of the dynamics of both the lower levels of the non-condensate band and 
of the quasiparticle levels in the condensate bands, which should provide the 
correct interpolation between the lower end of the non-condensate band and the 
upper levels of the condensate band.

On the other hand, when we are considering the dynamics of the condensate in 
equilibrium with the non-condensate---i.e. the equilibrium fluctuation dynamics 
of the condensate, the non-condensate distribution function {\em must} be the 
Bose-Einstein function, subject only to the levels being describable as 
the unperturbed trap levels.

\subsection{Stochastic master equation for diagonal matrix elements}
We can easily take the diagonal matrix elements of (\ref{simp1})
and derive the simple birth-death stochastic master equation for these
\begin{eqnarray}\Label{prac2}
\dot p(N) &=&   2NW^+(N-1)p(N-1)
\nonumber \\&&
 + 2(N+1)W^-(N+1)p(N+1)
\nonumber \\&& 
-\Big(2(N+1)W^+(N)+2NW^-(N)\Big)p(N).
\end{eqnarray}
This is of a very standard form from which we can derive the following.
\subsubsection{Stationary solution}
This is given by
\begin{eqnarray}\Label{prac3}
P_s(N) &\propto& \prod_{M=1}^N {W^+(M-1)\over W^-(M)}
\end{eqnarray}
and using (\ref{simp6}), this can be written
\begin{eqnarray} \Label{prac4}
P_s(N)&=&
\exp\left(\sum_{M=1}^N[\mu-\mu(M)]/k_{B}T\right)
\\ \Label{prac5}&\propto&
 \exp\left(\left[\mu N -E_0(N)\right]/k_{B}T\right),
\end{eqnarray}
and is the result expected from statistical mechanics.  We can make a Gaussian 
approximation to $P_s(N)$ by expanding about its maximum, which occurs at the 
value $ \bar N$ determined by $ \mu=\mu(\bar N) $, and the variance of this 
Gaussian is given by
\begin{eqnarray}\Label{prac6}
{\rm var}(N) &=& k_{B}T\bigg/{\partial \mu(\bar N) \over\partial\bar N }.
\end{eqnarray}
The Thomas-Fermi approximation to the condensate chemical potential is
$ \mu(\bar N) \propto{\bar N}^{2/5}$, and when this is valid it can be seen 
that
\begin{eqnarray}\Label{prac7}
{\rm var}(N) \propto {\bar N}^{3/5}
\end{eqnarray}
which can be sub-Poissonian for sufficiently large $ \bar N$, although this may 
not be achieved in practical cases. 
\subsubsection{Order parameter and off-diagonal long range order}
The correlation function of relevance is
\begin{eqnarray}\Label{ODLR1}
\langle \phi^\dagger({\bf x},t)\phi({\bf x}',t)\rangle_s &=&
\sum_N N P_s(N)\xi_N^*({\bf x})\xi_N({\bf x}')
\\ \Label{ODLR2}
&\approx&{\bar N}\xi_{\bar N}^*({\bf x})\xi_{\bar N}({\bf x}')
\end{eqnarray}
whenever $ {\rm var}(N)$ is very small compared to $ \bar N$, which is clearly 
the case for large enough $ \bar N$.  However it is perfectly clear that
$\langle \phi^\dagger({\bf x},t)\rangle =\langle\phi({\bf x}',t)\rangle_s =0$, 
so that off-diagonal long range order is achieved without breaking the phase 
symmetry of the Hamiltonian $ \phi({\bf x}) \to  \exp(i\theta) \phi({\bf x}) $.

Of course the order achieved here is clearly not ``long range'', since the 
ground state wavefunction $ \xi_N({\bf x})$ is only non-zero over a very small
distance interval, unless the trap is very broad, which would no longer be 
within the range of validity of the present treatment, which assumes distinctly 
spaced energy levels.
\subsubsection{Fokker-Planck equation}
For sufficiently large $ N$ we can make a Kramers-Moyal expansion \cite{SM,vK} 
to the stochastic equation (\ref{prac2}), yielding the Fokker-Planck equation
\begin{eqnarray}\Label{prac8}
\dot p(N,t) &=& 2{\partial \over\partial N }
\bigg\{\left[NW^-(N)-(N+1)W^+(N)\right]p(N,t)
\nonumber \\ &&
             +{\partial \over\partial N }\left[NW^+(N)p(N,t)\right]\bigg\}.
\end{eqnarray}
This is equivalent to the stochastic differential equation (in the Ito form)
\cite{SM,vK}
\begin{eqnarray}\Label{prac9a}
dN &=& 2[(N+1)W^+(N)-NW^-(N)]\,dt 
\nonumber\\&&
+ 2\sqrt{NW^+(N)}\,dW(t).
\end{eqnarray}
Assuming once more small fluctuations, we can linearize about the mean 
stationary value $ \bar N$ by writing $ N=\bar N +n$, and use the relationship 
between the forward and backward rates (\ref{simp6}) to obtain the linear 
stochastic differential equation
\begin{eqnarray}\Label{prac9b}
dn &=& -k n \,dt + \sqrt{D}\,dW(t)
\end{eqnarray}
in which
\begin{eqnarray}\Label{prac10}
k&=& 2\bar N W^+(\bar N){1\over k_{B}T}{\partial\mu(\bar N) \over\partial\bar N 
}
\\ \Label{prac11}
D &=& 4\bar N W^+(\bar N).
\end{eqnarray}
The characteristic relaxation time of the number fluctuations is thus
given by $ 1/k$ for small fluctuations about the equilibrium value.

\subsubsection{The growth equation}
The rate equation arising from (\ref{prac2},\ref{prac9a}) for the mean number 
is obtained by neglecting all fluctuations and takes the form
\begin{eqnarray}\Label{simp699}
\dot N&=& 2[(N+1)W^+(N) -NW^-(N)],
\end{eqnarray}
and when the thermal form for relation between the forward and backward rates 
(\ref{simp6}) is taken, this becomes
\begin{eqnarray}\Label{simp7}
\dot N &=& 2W^+(N)\left\{\left(1-e^{(\mu(N)-\mu)/k_{B}T}\right)N +1\right\}.
\end{eqnarray}
This equation combines aspects of laser theory and chemical thermodynamics 
into one simple statement.  The picture is analogous to that of two chemical 
species, ``condensate'' with chemical potential $ \mu(N)$, and ``vapor'' with
chemical potential $ \mu$, which come to equilibrium when the number of atoms 
in the condensate $ \bar N$ satisfies the requirement
\begin{eqnarray}\Label{equil}
\bar N = 1\bigg / \left(e^{[\mu(\bar N) -\mu]/k_{B}T}-1\right).
\end{eqnarray}
Since it is assumed that $ \bar N \gg 1$, this means that 
$ \mu(\bar N) = \mu$, to order $ 1/\bar N$.  This mean corresponds to the 
modal value, $ N_{\rm mod}$, of $ P_s(N)$ determined by $ W^+(N_{\rm mod}-1) 
=W^-(N_{\rm mod}) $ which 
from (\ref{simp6}) requires $ \mu(N_{\rm mod}) = \mu$.

We will call equation (\ref{simp7}) the {\em growth equation}; it can easily be 
integrated numerically, and the results have already been published in 
\cite{BosGro}.  The characteristic behavior is best illustrated when we start 
with the initial condition with $ N=0$, and assume that the vapor has 
sufficient atoms for us to neglect its depletion as atoms enter 
the condensate, so that $ \mu$ is assumed to be constant.  Under these 
conditions the initial growth of $ N$ is given by the ``spontaneous'' term in 
the growth equation (\ref{simp7}), i.e., $ \dot N = 2 W^+(N)$.  This gives a 
relatively slow growth, until $ N$ becomes large enough for the ``gain'' term 
proportional to $ N$ become significant---this term is positive if 
$\mu > \mu(0) $, and yields a dramatic net gain, giving a rapid rise in $ N$
which saturates as $ N$ approaches the stationary value $ N_s$, for which
$ \mu = \mu(N_s)$.
Rubidium


\subsection{Phase diffusion}
This section gives the simplest possible discussion of phase diffusion, for 
which we will consider 
the single-mode master equation in the form (\ref{simp1}).  The additional 
terms 
(\ref{comp9}) will have a significant effect on phase diffusion, but for 
simplicity are not included here.  A full treatment of phase diffusion will be 
given in \cite{QKIV}.

The phase decay is given by the 
equilibrium time-correlation function
$ G({\bf x},t,{\bf x}',t') \equiv
 \langle\phi^\dagger({\bf x},t)\phi({\bf x}',t')\rangle$, and we can evaluate 
this using standard techniques \cite{QNoise}
\begin{eqnarray}\Label{phase1}
 G({\bf x},t,{\bf x}',t') = \xi^*({\bf x})\xi({\bf x}')
{\rm Tr}\left\{A^\dagger\left\{e^{L(t-t')}A\rho_s\right\}\right\}
\end{eqnarray}
where $ L$ is the master equation evolution operator defined by writing 
(\ref{simp1}) in the form $ \dot \rho = L\rho$.
Notice that the operator expansion in terms of $ A$ is used since this 
preserves the basis state expansion; thus $ A^\dagger\rho_s$ can be written 
straightforwardly as an operator of the type (\ref{dens1}).

If we define
\begin{eqnarray}\Label{phase2}
\rho'(\tau) \equiv e^{L\tau}A\rho_s
\end{eqnarray}
then the object we need to compute is 
\begin{eqnarray}\Label{phase3}
{\rm Tr}\left\{A^\dagger\rho'(\tau)\right\} =
\sum_N \sqrt{N-1}\langle N-1 |\rho'(\tau)|N\rangle
\end{eqnarray}
subject to the initial condition
\begin{eqnarray}\Label{phase4}
{\rm Tr}\left\{A^\dagger\rho'(0)\right\} =
\sum_N {N}\langle N|\rho_s |N\rangle
\equiv \langle N \rangle_s
\end{eqnarray}
Let us first compute 
$g_N(\tau) \equiv \langle N-1 |\rho'(\tau)|N\rangle$
subject to the initial condition 
$ g_N(0) = \sqrt{N}\langle N |\rho_s |N\rangle$.
The master equation yields an equation of motion for $ g_N(\tau)$ in the form
\begin{eqnarray}\Label{phase5}
&& \dot g_N = {i\mu(N)\over\hbar}g_N
\nonumber \\ && +
2\sqrt{N(N-1)}W^+(N-1)g_{N-1}-(2N+1)W^+(N)g_N
\nonumber \\ && +
2\sqrt{N(N+1)}W^-(N+1)g_{N+1}-(2N-1)W^+(N)g_N.
\nonumber \\&&
\end{eqnarray}
We now assume that $ g_N$ is only significantly different from zero when $ N$ 
is large and carry out an expansion similar to the Kramers-Moyal expansion
\cite{SM,vK}.  We first set 
\begin{eqnarray}\Label{6}
2\sqrt{N(N\pm1)} \approx 2N \pm 1 - {1\over 4N}
\end{eqnarray}
and then make the Kramers-Moyal expansion to get
\begin{eqnarray}\Label{phase7}
\dot g_N &\approx &\left({i\mu(N)\over\hbar}- {W^+(N)+W^-(N)\over4N}\right)g_N
\nonumber \\ && +
2{\partial \over\partial N }\left\{ [NW^-(N)-(N+1)W^+(N)]g_N\right\}
\nonumber \\ &&
+2{\partial^2 \over\partial N^2 }\left\{[NW^+(N)g_N\right\}.
\end{eqnarray}
The Fokker-Plank operator in the second two lines of this equation is  exactly 
the 
same as that if (\ref{prac8}), but the two terms on the left of the top line 
cause a major 
difference in the behavior of the solutions. The simplest way to elucidate the 
behavior of the solutions is to linearize the equations in much the same way 
as led to the approximate stochastic differential equations 
(\ref{prac9a}--\ref{prac11}).  This means that we write approximately
\begin{eqnarray}\Label{phase8}
\mu(\bar N +n) &\approx & 
\mu(\bar N) + n{\partial \mu(\bar N) \over\partial \bar N}
\nonumber \\
&\equiv & \mu(\bar N) + \hbar\epsilon n,
\\  \Label{phase9}
{W^+(N)+W^-(N)\over4N} &\approx & {W^+(\bar N)+W^-(\bar N)\over4\bar N}
\nonumber \\
&\equiv & \kappa .
\end{eqnarray}
It is necessary to go to higher order in the first equation, (\ref{phase8}) 
because this term 
is imaginary, and may cause significant cancellation, which cannot happen for 
the second term, which is real.  The linearized equation then takes the form, 
written now in terms of the function $ f(n,\tau)$, defined by
$g(n+\bar N,\tau) \equiv \exp[(i\mu(\bar N)/\hbar -\kappa )\tau] f(n,\tau)$ of 
the variable $ n$
\begin{eqnarray}\Label{phase10}
\dot f(n) &=& i\epsilon n f(n) 
+ {\partial \over\partial n}\left\{k n f(n) 
+ {D\over 2}{\partial \over\partial n} f(n)\right\}.
\end{eqnarray}
This equation can be solved by the same method as used in \cite{SM} Sect.3.8.4.  
The initial condition follows from that for $ g( N)$, and, to the degree of 
accuracy being used in this linearized treatment, can be taken as
$ f(n,0) = \bar N P_s(\bar N + n)$.  This leads to the solution
for the characteristic function 
\begin{eqnarray}\Label{phase 11}
&& \tilde f(s,\tau) \equiv \int dn\,e^{ins}f(n,\tau)
\\ &&\quad =
\exp\left(-{Ds^2\over 4k} -{D\epsilon\over2k^3}\left(1-e^{-k\tau}\right)
-{D\epsilon^2\over 2k^2}\tau\right)
\end{eqnarray}
From the initial condition (\ref{phase4}) it follows that 
\begin{eqnarray}\Label{phase12}
&&{\rm Tr}\left\{A^\dagger\rho'(\tau)\right\}= e^{(i\mu(\bar N) -\kappa )\tau}
\bar N \int dn f(n,\tau)
\\ &&\quad= e^{(i\mu(\bar N)/\hbar -\kappa )\tau}\bar N\tilde f(0,\tau)
\\ &&\quad= e^{(i\mu(\bar N)/\hbar -\kappa )\tau}\bar N
\exp\left({D\epsilon^2\over2k^3}\left[1-k\tau-e^{-k\tau}\right]\right)
\end{eqnarray}
This equation shows the expected oscillatory behavior determined by the 
chemical potential $ \mu(\bar N)$, and three characteristic damping time 
constants.

\noindent (i): The time constant $ 1/\kappa$, which is the analogue of phase 
diffusion in a laser, and has its characteristic form of the noise divided by 
the number of particles.  We call the corresponding time constant
\begin{eqnarray}\Label{phase13}
\tau_1 \equiv {4\bar N\over W^+(\bar N)+W^-(\bar N)}
= {\pi \hbar^2\bar N \over2ma^2(k_B T)^2}
\end{eqnarray}

\noindent (ii): The behavior for $ kt \ll 1$ of the final exponential factor is 
of the Gaussian form, characteristic of inhomogeneous broadening,
$ \exp\left[-(\tau/\tau_2)^2/2\right]$, where
\begin{eqnarray}\Label{phase14}
\tau_2 \equiv 
\sqrt{{\hbar^2\over k_BT}
\left\{\partial\mu(\bar N) \over\partial\bar N \right\}^{-1}}
\end{eqnarray}

\noindent (iii): For $ kt\gg 1$, the behavior of this term becomes exponential
with a time constant
\begin{eqnarray}\Label{phase15}
\tau_3 = {2k^2\over D\epsilon^2}
\approx {8ma^2\bar N\over \pi \hbar}.
\end{eqnarray}
This time constant is independent of temperature, unlike the other two.


\subsection{Stochastic master equation including quasiparticles}
Let us now take into account the quasiparticle terms which arise from the terms 
(\ref{f9},\ref{f10a}) and develop a stochastic master equation for the 
occupation probabilities.  We can define an occupation probability
\begin{eqnarray}\Label{quasi1}
 p(N,{\bf n}) &=& \langle N,{\bf n}|\rho |N,{\bf n}\rangle
\end{eqnarray}
where $ {\bf n} =\{n_m\}$, the set of all quasiparticle 
occupation numbers.  The master equation terms (\ref{f9},\ref{f10a}) then give 
rise to  a stochastic master equation for these occupation probabilities
$ p(N,{\bf n})$, in the form 
\widetext
\begin{eqnarray}\Label{stochastic}
\dot p(N,{\bf n}) &=& 2NW^+(N-1)p(N-1,{\bf n})-2(N+1)W^+(N)p(N,{\bf n})
\nonumber \\  &+&
2(N+1)W^-(N+1)p(N+1,{\bf n})-2NW^-(N)p(N,{\bf n})
\nonumber \\  &+&\sum_m\{ 2n_mW_m^{++}(N-1)p(N-1,{\bf n}-{\bf e}_m)
                   -2(n_m+1)W_m^{++}(N)p(N,{\bf n})\}
\nonumber \\  &+&\sum_m\{ 2(n_m+1)W_m^{--}(N+1)p(N+1,{\bf n}+{\bf e}_m)
                   -2n_mW_m^{--}(N)p(N,{\bf n})\}
\nonumber \\  &+&\sum_m\{ 2(n_m+1)W_m^{+-}(N-1)p(N-1,{\bf n}+{\bf e}_m)
                   -2n_mW_m^{+-}(N)p(N,{\bf n})\}
\nonumber \\  &+&\sum_m\{ 2n_mW_m^{-+}(N+1)p(N+1,{\bf n}-{\bf e}_m)
                   -2(n_m+1)W_m^{-+}(N)p(N,{\bf n})\}
\end{eqnarray}
\narrowtext
Here $ {\bf e}_m = \{\dots 0,0,1,0,0,\dots \}$ has its only nonzero value
at the position corresponding to the index $ m$. 

There is no contribution to this equation from the terms
(\ref{comp9}), since these do not transfer particles between the vapor and the 
condensate, and we assume at this stage that the  $ 3\phi$ terms in 
(\ref{comp10},\ref{comp10a}) are negligible.

\subsubsection{Scattering of quasiparticles}
The stochastic master equation (\ref{stochastic}) is clearly incomplete, since 
it lacks the terms which would arise from the terms (\ref{comp9}), 
which represent the scattering of quasiparticles by atoms in the noncondensate 
band, thereby effecting a transfer of population between quasiparticle levels.  
The inclusion of these terms will not affect the stationary solutions, but 
could cause significant changes to time dependent quantities, such as 
condensate growth or time correlation functions.  Nevertheless we shall not 
pursue these aspects now, but will deal with them in a later publication.

One should also note that there can also be scattering of quasiparticles by 
each other, but these terms only appear if one takes the expansion in inverse 
powers of $ \sqrt{N}$ of the Hamiltonian $ H_0$ to the next two higher orders 
after those considered in (\ref{tb11}).  This aspect will also be left to a 
later publication

\subsubsection{Stationary solution}
Using similar reason to that in (\ref{thermal}) we can show that
\begin{eqnarray}\Label{stat1}
W_m^{++}(N+1) &=& e^{[\mu -\mu(N) -\epsilon_m]/k_{B}T}W_m^{--}(N)
\\ \Label{stat2}
W_m^{+-}(N+1) &=& e^{[\mu -\mu(N) +\epsilon_m]/k_{B}T}W_m^{-+}(N),
\end{eqnarray}
and from these it is straightforward to show that the stationary solution of 
the 
stochastic master equation (\ref{stochastic}) takes the expected form
\begin{eqnarray}\Label{stat3}
P_s(N,{\bf n}) &\,\propto \,&
\exp\left\{-{E_0(N) + \sum_m n_m\epsilon_m\over k_{B}T}\right\}.
\end{eqnarray}

\subsubsection{Deterministic equations of motion}
Assume factorization of all correlations, and neglecting 1 compared to $ N$, 
the deterministic equations of motion, analogous to the growth equation 
(\ref{simp7}), take the following form.  First define
\begin{eqnarray}\Label{mplus}
\dot n^+_m 
&\equiv& 2W^{++}_m(N)\left((1-e^{(\mu(N)-\mu+\epsilon_m)/k_{B}T})n_m+1\right)
\\ \Label{mminus}
\dot n^-_m 
&\equiv& 2W^{-+}_m(N)\left((1-e^{(-\mu(N)+\mu+\epsilon_m)/k_{B}T})n_m+1\right)
\end{eqnarray}
Then
\begin{eqnarray}\Label{dot1}
\dot n_m &=& \dot n^+_m +\dot n^-_m
\\ \dot N &=& 2W^+\left((1-e^{(\mu(N)-\mu)/k_{B}T})N+1\right)
\nonumber  \\&&
+\sum_m\left\{ \dot n^+_m -\dot n^-_m\right\}
\end{eqnarray}
Even though these equations are derived by neglecting all correlations, the 
stationary solution is given to the degree of approximation 
used---and large $ N$---by firstly making the condensate chemical potential 
match that of the vapor, which requires to lowest order
\begin{eqnarray}\Label{mplus2}
\mu(N_s) &=& \mu 
\end{eqnarray}
and then requiring  $ \dot n_m^\pm =0$, which then gives
\begin{eqnarray}
n_{m,s} &=& {1\over e^{\epsilon_m/k_{B}T}-1}
\end{eqnarray}
and these are the solutions expected from statistical mechanics.
The full time dependent solutions will be treated elsewhere, since they involve 
much more detail than we wish to give here.

\section{Conclusion}
\subsection{The scope of this and future work}
We have provided in this paper a full quantum kinetic description of a 
Bose-Einstein condensate in contact with a bath consisting of a vapor of 
non-condensed atoms at a well defined temperature and chemical potential.  The 
formulation up to Sect.\ref{Bog}, where the Bogoliubov approximation is 
introduced, is valid for any number of particles in the condensate, while the 
remainder of the paper, since it relies on the Bogoliubov approximation, is 
restricted to situations in which the number of particles in the condensate is 
rather large, perhaps of the order of a few hundreds.  However an extension of 
the methodology to deal with small numbers of particles in the condensate would 
be straightforward, and requires us simply to express the field operators in 
terms of particle creation and destruction operators, and rewrite 
(\ref{II4.11a}--\ref{II4.11f}) appropriately.

The key to the understanding of the condensate dynamics in this paper is the 
use of the particle number conserving Bogoliubov method \cite{trueBog}, which 
enables us to use the Bogoliubov spectrum as a function of the precise total 
number $ N $ of particles in the condensate band  (and {\em not\/ } merely the 
mean value $ \langle N\rangle$) , and further, to write precise expressions for 
the relevant transition matrix elements which appear in the master equation.  
The results of this procedure give very simple and easily understood equations 
of motion, especially when we; (a) ignore all quasiparticle effects; and
(b) include only the simplest of the terms which contribute to the equations of 
motion, those which correspond to gain and loss from the condensate as a result 
of collisions between particles in the non-condensate band. This will clearly 
be possible at low enough temperatures and with sufficiently large condensates.  
The resulting equations are very like those used in laser theory 
\cite{QNoise,Walls-Milburn,Haken}, and thus are able to be handled by similar 
means.  We have exhibited the basic theory of condensate number fluctuations; 
our treatment of the theory of phase fluctuations is simplified considerably, 
with  a full treatment to appear in another paper \cite{QKIV}, 
since these are measurable and of great interest, especially for the 
construction of an atom laser \cite{atom-lasers}, and thus merit a very full 
and careful treatment.  We will note here only that the methodology is 
straightforward, and very similar to that used for laser theory.

The growth equation, (\ref{simp7}), which represents in simplest terms the 
growth of the condensate from a vapor has already been reported \cite{BosGro}, 
and is in good general agreement with experiment.  However, the process of 
condensate growth spans a wide range of computational and theoretical regimes, 
and the growth equation cannot be valid in all of these, since there will be 
corrections arising from both the inclusion of quasiparticle effects as well as 
those which must be made because the Bogoliubov approximation cannot be valid 
in the regime of small condensate numbers.  The details of the modifications 
resulting from these effects, as well as those which arise because of depletion 
of the vapor, together with some detailed consideration of the experimental 
regimes will be reported in another paper \cite{QKVI}.

The depletion of the vapor in the non-condensate band is an issue which 
requires for its confident handling a full treatment of the coupled kinetics of 
the condensate and the non-condensate band, and this will be reported in 
another paper \cite{QKIV}, though the methodology will turn out mainly to 
depend on the use of equations of the Uehling-Uhlenbeck \cite{UU} kind for the 
non condensate, coupled to master equations, of the kind derived in this paper,
for the condensate.

\subsection{Validity of the approximations}
The simplifications and approximations made in this paper can be summarized as 
follows:
\begin{itemize}
\item[i)] In going from (\ref{H9}) to (\ref{met}) we have neglected the 
interaction between the condensate and the non-condensate bands in the kernel.  
This means that we assume that the condensate band has a spectrum which is not 
greatly changed by this interaction.  For condensates presently being studied, 
this is certainly experimentally valid for the lowest levels, since their 
behaviour is well described by the time-dependent Gross-Pitaevskii equation.  
However, to give a quantitative criterion for the validity of this 
approximation seems at this stage to be too ambitious.

\item[ii)] After the previous approximation, a Markov approximation is made.  
This amounts to requiring that the irreversible processes arising from the 
master equation are much slower than the reversible processes described by the 
separate evolution operators of the condensate and non-condensate bands.  For 
the non-condensate band, the arguments for its validity are the same as those 
in QKI, while for the condensate band, the evidence is experimental, that the 
damping measured is definitely weak compared to the timescale of oscillations, 
which can only be the consequence of the condensate band Hamiltonian $ H_0$.

\item[iii)] Another major approximation is the rotating wave or random phase 
approximation.  Again, it is difficult to give a quantitative measure of its 
validity, 
other than to say that it should be valid when the basic behavior of the 
noncondensed vapor should be describable as transitions between states which 
are well described by single particle energy levels.  All the exprimental 
evidence, including the very design of the experiments, seems to verify this.  
The basic premise of the random phase approximation is that the non-resonant 
terms are all of random signs, and effectively change sign frequently within 
times of interest, and it seems to us that the arguments for the validity of 
the random phase approximation are as good here as in similar situations in 
solid state physics.  However, one can only make a definite statement by 
comparison with experiment.
\end{itemize}
The question of validity can also be looked at from another point of view; 
namely, there will be traps, of the type illustrated in Fig.1(a), in which the 
separation between the dynamics of the condensate and non-condensate band is as 
good as one wishes, and for which the approximations employed will certainly be 
valid.  Even so, we believe that it is very likely that our approximations are 
already good for the Harmonic traps currently in use, but ultimately it seems 
to us that other kinds of trap may be both more useful, and more adapted to our 
kind of description.
\subsection{Relation to other work}
Apart from the work of Anglin \cite{Anglin}, previous work on the field of 
condensate growth \cite{KaganKinetics,Trento,Stoof91,Semikoz}
has focussed  on the  growth of a condensate in a spatially homogeneous system,
or an approximately spatially homogeneous system, 
rather than in a trapped system, and thus has not been applicable to the 
condensates presently being produced.
The major difference in point of view in this paper is the description of the 
condensate in terms of the trap energy levels for a given number of particles, 
which is treated here by the modified Bogoliubov method, and the clear division 
into the condensate and non-condensate bands.  It is these two innovations 
which make our treatment of condensate growth possible.

Our simplified master equation (\ref{simp1}) bears a 
very strong resemblance to that of Anglin \cite{Anglin}, who presented 
a model calculation based on a single level in a very tight potential 
well, centered inside a spherically symmetric bath of atoms moving 
essentially freely, apart from collisions.  Anglin's model can be viewed as the 
description of the most ideal trap imaginble, in which the small trap permits 
only one energy level, and the larger trap is actually almost infinite.  It is 
thus drastically simplified from the beginning, unlike ours, which treats 
all relevant features as exactly as possible.  The term in Anglin's 
model which gives pure phase diffusion is not present in 
(\ref{simp1}), but clearly arises from similar approximations applied 
to the term (\ref{comp9}).  Indeed if one takes the diagonal matrix 
elements of Anglin's master equation, one obtains an equation of the 
form (\ref{prac2}) in which $ W^+$ and $ W^-$ are related to each 
other by (\ref{simp6}), and the absolute value of $ W^+$ is given by 
multiplying our value (\ref{prac1}) (apart from the final term in 
curly brackets) by a factor of order of magnitude $ kT/E_b$, where $ 
E_b$ is the binding energy of Anglin's potential as defined in his 
paper.

The major assumption behind Anglin's derivation of the Markov approximation is 
that $ \exp(\mu/kT)$ is small.  This requires that $ \mu$ be negative, and that  
$-E_b < \mu $.  Thus the physical situation is rather different from what we 
consider---the vapor of uncondensed atoms is always essentially classical, and 
the condensation only occurs because of the depth of the attractive potential.  
The collisions which establish the condensate are of the same kind as ours, but
in Anglin's case a typical collision must yield one final atom in the 
condensate, whose negative energy $ \approx -E_b$ must be balanced by 
transferring the equivalent amount to the other atom, which must also take up 
all the momentum of the initial two vapor particles---this will involve 
typically particles with energy $ \approx E_b$.  If $ kT \ll E_b $
this will be a rather small fraction of the total number of particles, and 
hence the extra factor $ kT/E_b$ seems quite understandable.  

Thus we conclude that Anglin's model is consistent with our analysis.

\subsection{The broken symmetry picture}
Our picture does appear to be at variance with the description of the 
condensate in terms of  ``broken gauge symmetry''or ``Bose broken symmetry'', 
which has been customary for many years \cite{Leggett,Sols,Griffin}.  It 
appears to us that these concepts are in fact not helpful in the description of 
the condensates which are produced in current experiments, and indeed we doubt 
whether they have any relevance to the physics.  The stationary states of the 
system are indeed in our formulation simple statistical mixtures of number 
states of $ N $, the total number of particles in the condensate band.  The 
correlation functions $ \langle \phi^\dagger({\bf x})\phi({\bf x}')\rangle$
however do exhibit the mandatory property of ``off diagonal long range order'', 
even though the stationary averages 
$ \langle \phi^\dagger({\bf x})\rangle$, $\langle\phi({\bf x}')\rangle$ are
zero.  This happens in laser theory as well, where the existence of a well 
defined phase of the optical field is the result of a rather long phase 
diffusion time compared with the other relaxation times---we will demonstrate 
the analogous property for the Bose condensate in \cite{QKIV}, though the 
essential points have already been foreshadowed by several authors
\cite{Javanainen96,Cirac,Molmer,Castin-Dalibard}, who have shown how the 
measurement of the 
relative phase between two condensates introduces a relative coherence between 
the two condensates, which remains in existence on the time scale of phase 
diffusion.  In a ferromagnet the concept of spontaneous symmetry breaking 
acquires its validity from the extremely long diffusion time for the magnetic 
moment of a sample of the ferromagnet---over a sufficiently long time the 
average magnetic moment would be zero, but this time is much longer than the 
lifetime of the average experimenter, who therefore does not see things this 
way.  In the end it is a matter of taste where we draw the line between 
dephasing time which is very long and one which is essentially infinite---it 
depends on the relevant timescales of the experiment, including ultimately the
timescale of the patience of the experimenter.

\acknowledgements
We would like to acknowledge discussions with R. J. Ballagh, M. J. Davis, J. I. 
Cirac, D. Jaksch, J. Anglin and G. Shlyapnikov.

This work was supported by the Marsden Fund under contract number
PVT-603, and by \"Osterreichische Fonds zur F\"orderung der 
wissenschaftlichen Forschung.
\appendix
\section{Estimation of the size of the condensate band}\label{app1}
In this appendix we will use the Bogoliubov method as described in 
Sect.\ref{Bog} to get an estimate of the energy $ E_R$ above which we may 
neglect the effect of the condensate on the trap energy levels.
\subsubsection{Thomas-Fermi approximation}
We use the Thomas-Fermi approximation to the condensate wavefunction in which 
we neglect the Laplacian term in the time independent Gross-Pitaevskii 
equation, giving the approximate solution 
\begin{eqnarray}\Label{tf1}
\xi_N({\bf x}) &=& \sqrt{\mu(N)-V_T({\bf x})\over N u}.
\end{eqnarray}
For a harmonic potential
\begin{eqnarray}\Label{tf2}
V_T({\bf x}) &=& ax^2 + by^2 + cz^2
\end{eqnarray}
normalization requires
\begin{eqnarray}\Label{tf3}
\mu(N) &=& \left({15Nu\sqrt{abc}\over 8\pi}\right)^{2/5}
\end{eqnarray}
where
\begin{eqnarray}\Label{tf301}
a=m\omega_x^2/2;
\quad
b=m\omega_y^2/2;
\quad
c=m\omega_z^2/2;
\end{eqnarray}

\subsubsection{Quasiparticle spectrum}
Use the definition (\ref{t5c}) of $ {\cal H}_3$, and substitute (\ref{tf1})
to get 
\begin{eqnarray}\Label{tf4}
{\cal H}_3 &=&
\int d^3{\bf x}\bigg\{
-{\hbar^2\over 2m}\chi^\dagger({\bf x})\nabla^2\chi({\bf x})+
\chi^\dagger({\bf x})V_{TF}({\bf x})\chi({\bf x})\bigg\}
\nonumber \\
&+&  {\tilde u\over 2}\int\limits_{V({\bf x})<\mu(N)}{\!\!\!\!\!\!\!}
 d^3{\bf x}
:\!\big(\chi^\dagger({\bf x}) +\chi({\bf x})\big)^2\!:
\left(\mu(N) -V_T({\bf x})\over Nu \right)
\nonumber \\ &+&{\rm const.}
\end{eqnarray}
with
\begin{eqnarray}\Label{tf5}
V_{TF}({\bf x})&=& {\rm max}\{V_T({\bf x}) - \mu(N),0\}
\end{eqnarray}
We can then write
\begin{eqnarray}\Label{tf6}
{\cal H}_3
&=& \sum_q \bar E_q a^\dagger_qa_q 
+ \sum_{k,q}G_{kq}(2a^\dagger_qa_k +a^\dagger_qa^\dagger_k +a_qa_k )
\end{eqnarray}
where 
\begin{eqnarray}\Label{tf7}
G_{kq} &=& \int_{V({\bf x})<\mu(N)} d^3{\bf x}\,
\xi_q({\bf x})\xi_k({\bf x})\Big(\mu(N) -V_T({\bf x})\Big),
\end{eqnarray}
$ \bar E_q$ are the energy levels of $ V_{TF}$, and $ \xi_q({\bf x})$ are the 
corresponding eigenfunctions, which have been chosen as real.
\subsubsection{Approximate evaluation of integrals}
For high lying excitations, we expect that the smooth nature of the ground 
state wavefunction will mean that essentially  $ G_{kq}\approx \delta_{kq}g_q 
$, so that the individual modes are diagonalized by a {\em squeezing 
transformation}
\begin{eqnarray}\Label{tf8}
a_q = c_qA_q - s_q A^\dagger_q
\end{eqnarray}
with, for small $ g_q/\bar E_q$,
\begin{eqnarray}\Label{tf9}
s_q &\approx& g_q/\bar E_q
\\ \Label{tf10}
c_q &\approx & 1-s_q^2
\end{eqnarray}
To estimate $ g_q$, we approximate the wavefunction $ \xi_q$ by a WKB 
approximation to an appropriate three dimensional harmonic oscillator 
wavefunction (obtained
by neglecting the flattening in the potential caused by the condensate, but not 
the chemical potential shift); this means 
we can approximate by smoothing over the oscillations, and then neglecting the 
$ {\bf x}$ dependence, to get
\begin{eqnarray}\Label{tf11}
\xi_q({\bf x})^2 \approx D_{n_1,n_2,n_3}\equiv
P_x(n_1,0) P_y(n_2,0) P_z(n_3,0)
\end{eqnarray}
with
\begin{eqnarray}\Label{tf12}
P_x(n_1,x) = {1\over  \pi}
\left[{(2n_1+1)\hbar\over m\omega_x}-x^2\right]^{-1/2}
\end{eqnarray}
with similar equations for the other terms. This means that (letting
$ q\to \{n_1,n_2,n_3\}$,
\begin{eqnarray}\Label{tf13}
g_{n_1,n_2,n_3}&= & D_{n_1,n_2,n_3}
         \int_{V({\bf x})<\mu(N)} d^3{\bf x}\Big(\mu(N) -V_T({\bf x})
\Big)
\nonumber\\
\\ \Label{tf14}
&=& {16\mu(N)\sqrt{2}\over15\pi^2}
\sqrt{\mu(N)^3\over\hbar^3\prod_{i=1}^3(2n_i+1)\omega_i}
\\ \Label{tf16}
\bar E_{n_1,n_2,n_3} &\approx&
{ \hbar\over 2}
\sum_{i=1}^3 (2n_i+1)\omega_i-\mu(N)
\end{eqnarray}
\subsubsection{Numerical comparisons}
For the case of the TOP trap of JILA using Rubidium, we find that the 
ratio $g_{n_1,n_2,n_3}/ \bar E_{n_1,n_2,n_3}$ can be chosen to be less than 
about 0.03 for all $ {n_1,n_2,n_3}$ on a fixed energy surface provided
$ \bar E/\mu(N)$ is more than about 2.4 at $ N=1,000$ rising to 3.4 when
$ N=1,000,000$.

\section{The local equilibrium approximation for $ R_{NC}$}\Label{evaluations}
We want to calculate 
\begin{eqnarray}\label{eval1}
W^{+} &=& {u^2\over (2\pi)^5\hbar^2}
\int d^3{\bf K}_1\int d^3{\bf K}_1\int d^3{\bf K}_1\int d^3{\bf k}
\nonumber\\&&\times\delta({\bf K}_1 +{\bf K}_2-{\bf K}_3-{\bf k})
\nonumber \\&&\times\delta\left(\Delta_{123}(0)-\mu(N)/\hbar\right)
F({\bf K_1},0)F({\bf K_2},0)
\nonumber \\&&\times
[1+F({\bf K_3},0)] |\tilde\xi({\bf k}) |^2.
\end{eqnarray}
We assume
\begin{itemize}
\item[i)] That we can use the Maxwell-Boltzmann form
\begin{eqnarray}\label{eval1a}
F({\bf K},0)\approx\exp\left(\mu- {\hbar^2{\bf K}^2/2m}\over k_{B}T\right)
\end{eqnarray}

\item[i)] That $ F({\bf K_3},0)$ is negligible compared to 1 in the region of 
interest.
\item[ii)] That the range of $ {\bf k}$ in which $ |\tilde\xi({\bf k})|^2$
is non-negligible is very small.  This means we can drop all the remaining $ {
\bf k}$-dependence, and integrate $ |\tilde\xi({\bf k})|^2$ to give 1.
\end{itemize}
We then carry out the $ {\bf K}_3$ integration to get
\begin{eqnarray}\label{eval2}
W^{+}&=&  {u^2\over (2\pi)^5\hbar^2}\int d^3{\bf K}_1\int d^3{\bf K}_1
\nonumber \\ &&\times
\delta\left(\left[{{\hbar^2\over 2m}
\left({\bf K}_1^2+{\bf K}_2^2-({\bf K}_1+{\bf K}_2)^2\right)}-\mu(N)\right]/
\hbar\right)
\nonumber \\&&\times
\exp\left(2\mu -{\hbar^2}\left({\bf K}_1^2+{\bf K}_2^2\right)/2m\over k_{B}T
\right)
\end{eqnarray}
From this point no further approximations need to be made.  By changing 
variable to 
$ {\bf Q} = {\bf K}_1+{\bf K}_2$, 
$ {\bf q} = {\bf K}_1-{\bf K}_2$ the integral becomes spherically symmetric in  
both of the variables, and the $ {\bf q}$ integral can be carried out using the 
delta 
function, and the integral finally reduced to
\begin{eqnarray}\label{eval3}
W^+&=&{u^2\over (2\pi)^5\hbar^2}{2\pi^2m\over \hbar}
\left(4m\mu(N)\over\hbar^2\right)^2
\exp\left(2\mu-\mu(N)\over k_{B}T\right)
\nonumber \\&&\times
\int_0^\infty dt\, t^{1/2}(1+t)^{1/2}
\exp\left(-{2\mu(N)t\over k T}\right),
\end{eqnarray}
Using \cite{Abramowitz-Stegun}, formula 13.2.5, the integral can be reduced to 
a 
form
involving the confluent hypergeometric function $ U(3/2,3,z)$, which itself by 
formula 13.6.21 can be reduced to an expression involving the modified Bessel 
function $ K_1(z)$, which gives the final result
\begin{eqnarray}\label{eval4}
W^+(N) &=& {4m (ak_{B}T)^2 \over \pi \hbar^3}e^{2\mu/k_{B}T}
\left\{{\mu_N\over k_{B}T}K_1\left({\mu_N\over k_{B}T}\right)\right\},
\end{eqnarray}
as in (\ref{prac1}).


\begin{figure}\Label{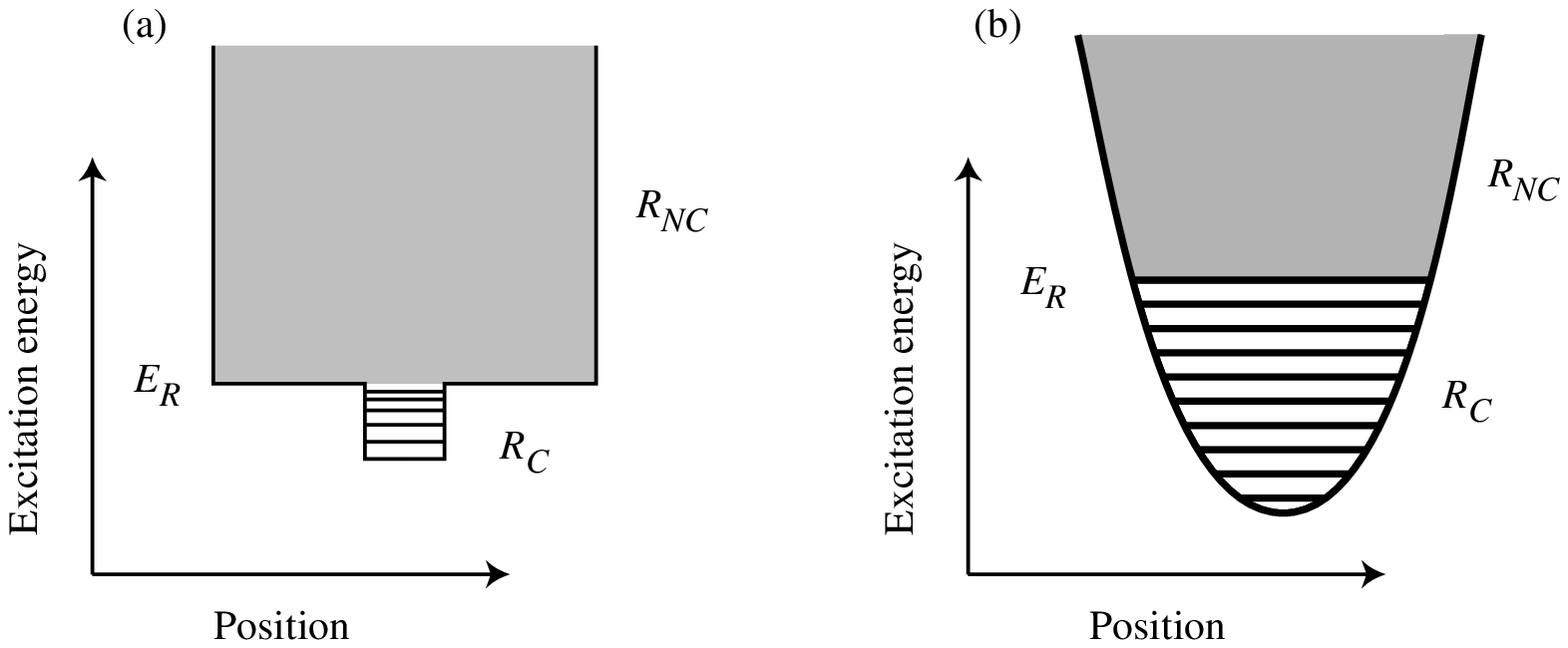}
\epsfig{file=fig1.eps, width=8cm}
\caption{
The condensate and non-condensate bands for (a) an ideal trap 
and (b) a harmonic trap.}
\end{figure}

\begin{figure}\Label{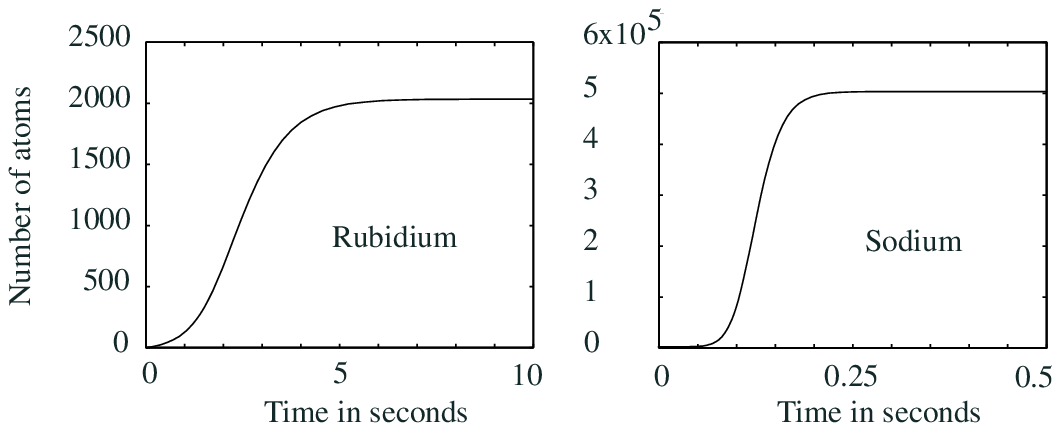}
\epsfig{file=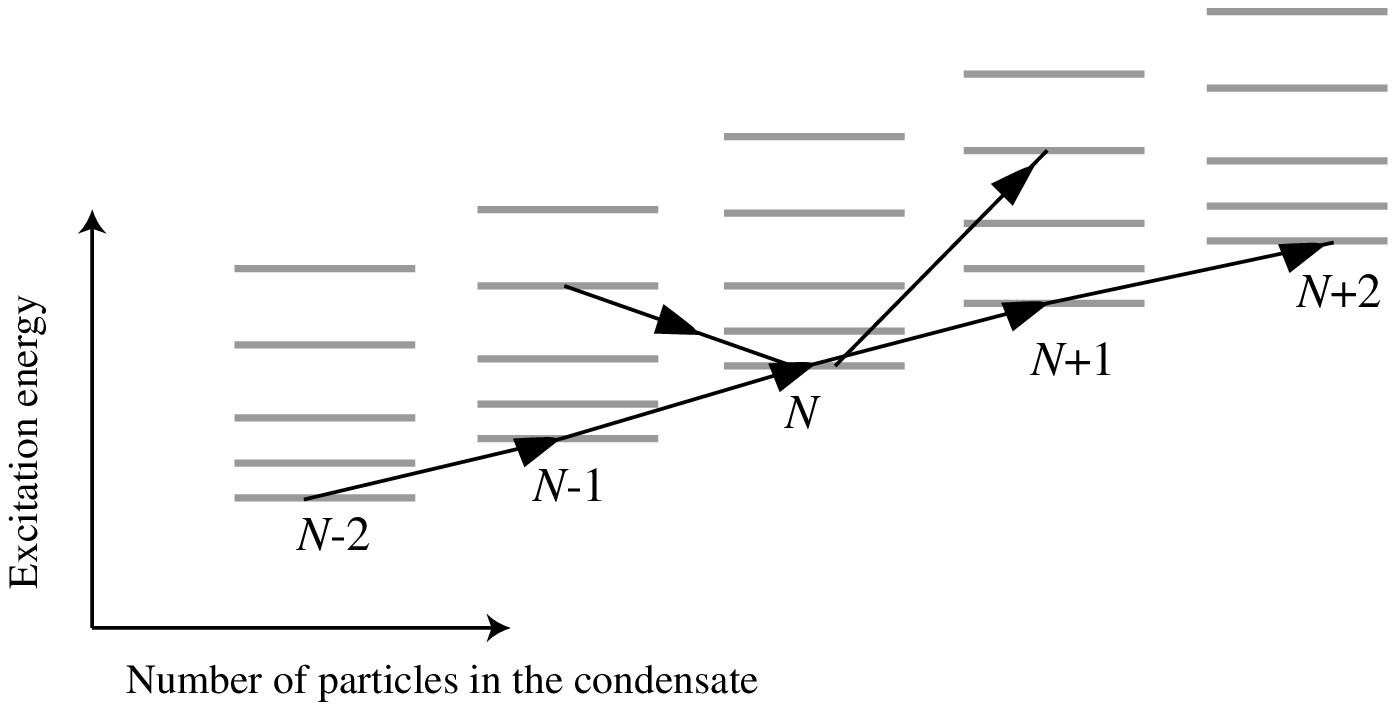, width=8cm}
\caption{
Transitions in terms of quasiparticles}
\end{figure}

\begin{figure}\Label{fig4}
\epsfig{file=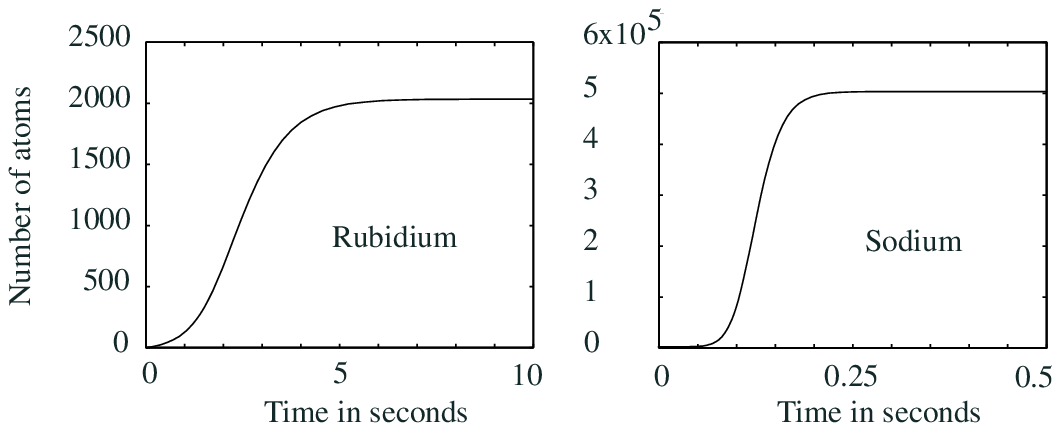, width=8cm}
\caption{
Condensate growth as given by the growth equation for (a) 
and (b) Sodium. Scattering lengths are $ 5.71\,{\rm nm}$ and 
$ 2.75\,{\rm nm}$, respectively.}
\end{figure}

\end{document}